\pdfoutput=1

\documentclass[twocolumn,apj]{emulateapj}

\shorttitle{First Detection of Chromospheric Magnetic Field Changes During an X1-Flare}

\usepackage{color}

\begin{document}

\title{First Detection of Chromospheric Magnetic Field Changes During an X1-Flare}

\author{Lucia Kleint\altaffilmark{1}}
\altaffiltext{1}{University of Applied Sciences and Arts Northwestern Switzerland, Bahnhofstrasse 6, 5210 Windisch, Switzerland}

\begin{abstract}
Stepwise changes of the photospheric magnetic field, which often becomes more horizontal, have been observed during many flares. Previous interpretations include coronal loops that contract and it has been speculated that such jerks could be responsible for sunquakes. Here we report the detection of stepwise chromospheric line-of-sight magnetic field (B$_{\rm LOS}$) changes obtained through spectropolarimetry of \ion{Ca}{2} 8542 \AA\ with DST/IBIS during the X1-flare SOL20140329T17:48. They are stronger ($<$640 Mx cm$^{-2}$) and appear in larger areas than their photospheric counterparts ($<$320 Mx cm$^{-2}$). The absolute value of B$_{\rm LOS}$ more often decreases than increases. Photospheric changes are predominantly located near a polarity inversion line, chromospheric changes near footpoints of loops. The locations of changes are near, but not exactly co-spatial to hard X-ray (HXR) emission and neither to enhanced continuum emission, nor a small sunquake. Enhanced chromospheric and coronal emission is observed in nearly all locations that exhibit changes of B$_{\rm LOS}$, but the emission also occurs in many locations without any B$_{\rm LOS}$ changes. Photospheric and chromospheric changes of B$_{\rm LOS}$ show differences in timing, sign, and size and seem independent of each other. A simple model of contracting loops yields changes of opposite sign to those observed. An explanation for this discrepancy could be increasing loop sizes or loops that untwist in a certain direction during the flare. It is yet unclear which processes are responsible for the observed changes, their timing, size and locations, especially considering the incoherence between photosphere and chromosphere.
 \end{abstract}

\section{Introduction}

Energy stored in the coronal magnetic field is released through the process of reconnection during flares. It is believed that this energy is the source for flare-related phenomena, such as particle acceleration, coronal mass ejections, heating and increased radiation throughout the spectrum, mass motions, changes in the magnetic field configuration, and sunquakes. The linkage between these processes however is not fully understood, for example it is yet unclear which processes cause the magnetic field in the lower solar atmosphere to change.

Permanent and often stepwise changes of photospheric line-of-sight magnetic fields, shear, and energies have been reported for many flares \citep[e.g.,][]{wang1992,wangetal1994,kosovichevzharkova1999, cameronsammis1999, kosovichevzharkova2001,sudolharvey2005}. Statistical analyses have found that B$_{\rm LOS}$ changes up to 450 G and that powerful flares exhibit stronger changes, with medians of 82 G and 54 G for X- and M-flares, respectively \citep{petriesudol2010}. The changes occur within minutes. They also seem to be accompanied by chromospheric emission, as seen in AIA 1600, which has been observed to start a few minutes earlier than the field change \citep{sudolharvey2005, johnstoneetal2012}. The opposite is not true, many UV brightening do not show a corresponding field change. The changes were reported to be co-spatial with the hard X-ray (HXR) emission during the early impulsive phase, but starting earlier than the HXR emission \citep{burtsevaetal2015}. 

Recent models predict a scenario of ``coronal implosions'' during which coronal loops contract and become more horizontal after flares \citep{hudson2000, hudsonetal2008, fisheretal2012}. These models are compatible to the findings of \citet{petriesudol2010} that stronger line-of-sight changes are observed near the solar limb and also to direct vector magnetic field measurements \citep[e.g.,][]{wangliu2010,wangliuwang2012}. The differences in timing between the different observables and the mechanism of energy transport from the corona to the lower solar atmosphere to drive the magnetic field changes - possibly by beams or by Alfv\'enic waves \citep{fletcherhudson2008} - are not yet understood. 

Changes in chromospheric magnetic fields during flares have been suspected, based on the changes of chromospheric features (e.g.\,fibrils, filaments) \citep{bruzek1975,zirin1983,labonte1987}. But no observation to date employed chromospheric polarimetry to unambiguously trace the magnetic field changes. Former chromospheric polarization measurements mostly focused on linear polarization and the search for ``impact'' polarization \citep[e.g.,][]{henouxetal1990,biandaetal2005,judgeetal2015}. In this paper, we report the first direct observations of chromospheric magnetic field changes that were observed through spectropolarimetry of the \ion{Ca}{2} 8542 \AA\ line during an X-class flare and we compare them to the photospheric changes of the same flare.

\section{Observations and data reduction}
\label{obs}

The X1 flare SOL20140329T17:48 \citep{kleintetal2015} occurred in AR 12017 at a heliocentric angle $\mu$=0.8. The flare was a consequence of a filament eruption, which started around 17:30 UT. The impulsive phase, as defined by the appearance of HXR emission above 25 keV, lasted from $\sim$17:45 - 17:49 UT.

To search for \textit{photospheric} changes of B$_{\rm LOS}$, data from the Helioseismic and Magnetic Imager (HMI) onboard the Solar Dynamics Observatory (SDO) \citep{scherrerhmi2012} were used. HMI  records six wavelength points with a filter passband of 76 m\AA\ in the \ion{Fe}{1} 6173 \AA\ line at a cadence of 45 s. The HMI plate scale is 0.5 arcsec pixel$^{-1}$. We use the series \textit{hmi.M\_45s\_nrt}, which interpolates linearly over three temporal intervals, instead of the interpolation with a sinc function over five intervals in the regular (non-nrt) data series. The algorithm used for HMI data to calculate B$_{\rm LOS}$ from the polarimetric images at these six wavelength points has a known scaling difference to the magnetic fields retrieved from the inversions and underestimates the actual field strength for strong fields \citep{hoeksemaetal2014}. We used data from 17:30-18:00 UT, removed the solar rotation, and aligned all images.

\begin{figure} 
  \centering 
   \includegraphics[width=.47\textwidth]{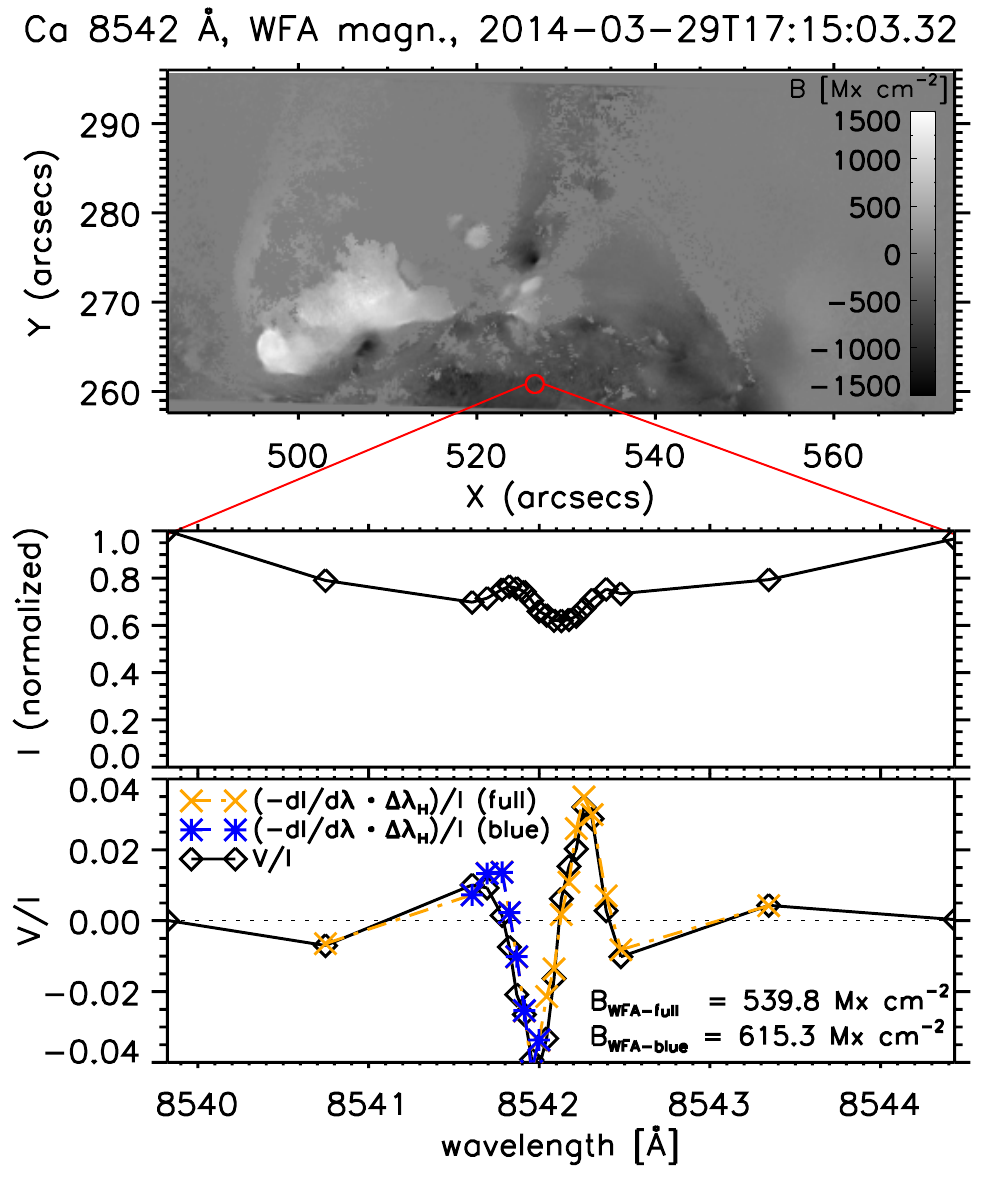}
   \caption{{Top: Chromospheric magnetogram. The (normalized) Stokes I profile of an example pixel (red circle, selected because it exhibits a significant change of the magnetic field of $\Delta$B$_{\rm LOS} = -220 $ Mx cm$^{-2}$, see Sect.~\ref{sec3}) is shown in the middle panel with the diamonds indicating the observed wavelengths. The bottom panel shows $V/I$ (black) and the two WFA fits obtained from the derivative of Stokes $I$ for the blue wing (blue) and the full profile (orange). Both fit well and the resulting magnetic field values are given on the bottom right.}}
           \label{exwfa}
  \end{figure}

To search for \textit{chromospheric} changes of B$_{\rm LOS}$, the \ion{Ca}{2} 8542 \AA\ line recorded by 
the Interferometric Bidimensional Spectropolarimeter (IBIS) at the ground-based Dunn Solar Telescope (DST) \citep{cavallini2006,reardoncavallini2008} was used. IBIS is a dual Fabry-Perot interferometer in a collimated beam. During our observations, three spectral lines were scanned in sequence (\ion{Ca}{2} 8542 \AA, H$\alpha$ 6563 \AA, and \ion{Fe}{1} 6302 \AA). Two different observing programs were run during the flare, resulting in a cadence of 56 s and a Ca line scan time of 18 s before 17:48 UT and a cadence of 37 s and a Ca line scan time of 15.5 s afterwards. The \ion{Ca}{2} 8542~\AA\ line was scanned with 21 wavelength points from 8539.8 to 8544.4~\AA\ with different step sizes (0.044 \AA\ near the line core to 1 \AA\ in the line wings, see Fig.~\ref{exwfa}). We use data from 16:45 - 18:23 UT for this study.

The IBIS data reduction includes a correction for dark current, gain, prefilter transmission, a destretch for seeing, the removal of the blueshift of the transmission profile across the FOV because of the collimated mount, and a polarization calibration \citep[for more details, see e.g.][]{kleint2012}. The noise in the polarization images is estimated at 1\% of $I$, which is about the signal in $Q$ and $U$. Therefore, only $V$ is used, i.e. only line-of-sight magnetic fields, and we restrict our analysis to pixels with max($V) \ge 2\%$.

We determine the chromospheric B$_{\rm LOS}$ from the weak-field approximation \citep[WFA; e.g.,][]{landi1992,harvey2012}. First, the data were binned by 2 (to 0.2 arcsec pixel$^{-1}$) to minimize influence from small misalignments. Then we applied the WFA separately to the whole profile (8540.7--8543.3 \AA) and to the blue wing (8541.6--8542.0 \AA), giving us two values for B$_{\rm LOS}$ and a method to check if the WFA is reliable or if the profile is irregular (see example in Fig.~\ref{exwfa}). The WFA relates Stokes $V$ to the derivative of $I$:
\begin{equation}\label{eq2}
V(\lambda) = -\Delta\lambda_H \cos\theta  \frac{d I(\lambda)}{d \lambda}
\end{equation}
where $\theta$ is the angle between the line-of-sight and the direction of $\vec B$, and $I(\lambda)$ is the intensity of the unsplit profile, which under the WFA is assumed to be close enough to the observed, magnetic profile. The Zeeman splitting is
\begin{equation}\label{zeemanspl}
 \Delta\lambda_H = \frac{e}{4\pi m_e c} B \lambda_0^2 \ g_{\rm eff} = 4.6686 \cdot 10^{-13} B \lambda_0^2 \ g_{\rm eff} 
\end{equation}
with the effective Land\'e factor $g_{\rm eff}$, the central wavelength $\lambda_0$ of the spectral line and units of \AA\ for wavelengths and G for the magnetic field strength. For \ion{Ca}{2} 8542 \AA, g$_{\rm eff}=$1.1. As a reminder for potentially unresolved fields and for the different resolutions of HMI and IBIS, we will use units of flux density (Mx cm$^{-2}$) instead of the field strength (G).

 \begin{figure*} 
  \centering 
   \includegraphics[width=.8\textwidth]{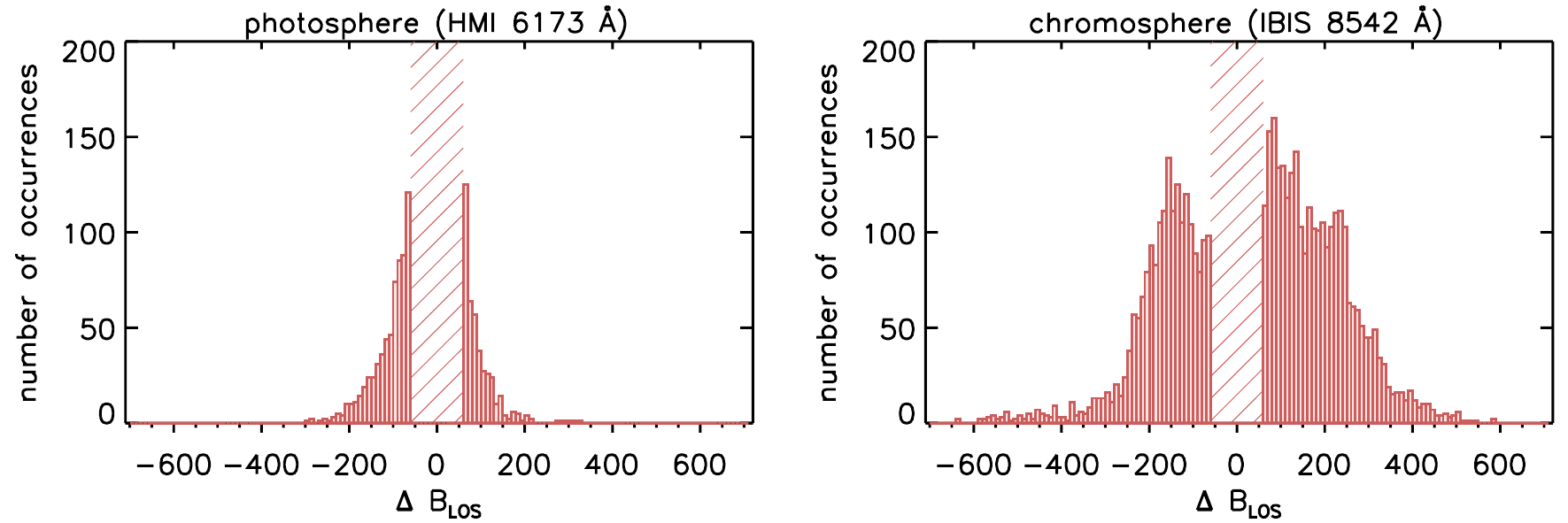}
   \caption{Histogram of LOS magnetic field changes in the photosphere (left) and in the chromosphere (right). The dashed area indicates changes below the noise level. Note that the numbers of changes differ also because of different spatial resolutions of HMI and IBIS. The chromospheric changes are larger and more asymmetric than those of the photosphere.}
        \label{histogram}
  \end{figure*}

\section{Determining the Magnetic Field Changes}\label{sec3}

To determine whether B$_{\rm LOS}$ in a given pixel exhibits a stepwise change, we fitted the same  function as \citet{sudolharvey2005}:
\begin{equation}\label{eq1}
B(t) = a + bt + c \left\{ 1  + \frac{2}{\pi} \tan^{-1} [n(t-t_0)] \right\},
\end{equation}
where $a$ and $b$ describe the background field evolution, $c$ is half of the amplitude of the magnetic field change, $1/n$ is the timescale and $t_0$ the midpoint of the change.  

Because the HMI algorithm fails at flare maximum due to the varying shape of Stokes $I$, we excluded those points (up to 5 points, depending on pixel) from fitting. We excluded bad fits of photospheric data with the following selection rules: the magnetic field change must be completed during our data series, the jump cannot be more than 1000~Mx~cm$^{-2}$ and a linear fit must not give smaller residuals than our stepwise fit. The remaining changes and all excluded changes above 50~Mx cm$^{-2}$ were then checked manually, because no mathematical algorithm matches the eye's ability to pick up the desired stepwise changes. This led to a final sample of $\sim$1100 pixels. As a crosscheck, we also fitted Eq.~\ref{eq1} to a time interval without any flares and while some pixels exhibit stepwise changes even then, there were more than five times fewer pixels that showed a 100 Mx cm$^{-2}$ change and in general, the step sizes were significantly lower during non-flaring times.

For \ion{Ca}{2}, the restriction of max($V) \ge 2\%$ led to a variable number of points to be fitted in the temporal data series of each pixel. Bad fits were excluded with the following conditions: 1) c $\le 0.15$ indicated sinusoidal and not stepwise variations of B$_{\rm LOS}$. 2) jump cannot be more than 1000 Mx cm$^{-2}$. 3) jumps before 17:32 and after 17:55 UT were deemed not to be related to the flare. 4) pixels with fewer than 15 data points before and 15 data points after the flare with a sufficient S/N were omitted. 5) pixels with fewer than 5 usable points at flare time $\pm$ 10 minutes were omitted. 6) pixels where a linear fit gave smaller residuals than our model were omitted. 7) pixels where the step size from blue wing WFA and the full WFA differed by more than 150 Mx cm$^{-2}$ were omitted, because they indicated irregular profiles where the WFA cannot be applied. All selected pixels and previously rejected pixels with a step size above 60 Mx cm$^{-2}$ were verified by eye and those without a clear stepwise change rejected, which led to a final sample of $\sim$4600 pixels. 

\begin{figure*} 
  \centering 
   \includegraphics[width=.48\textwidth]{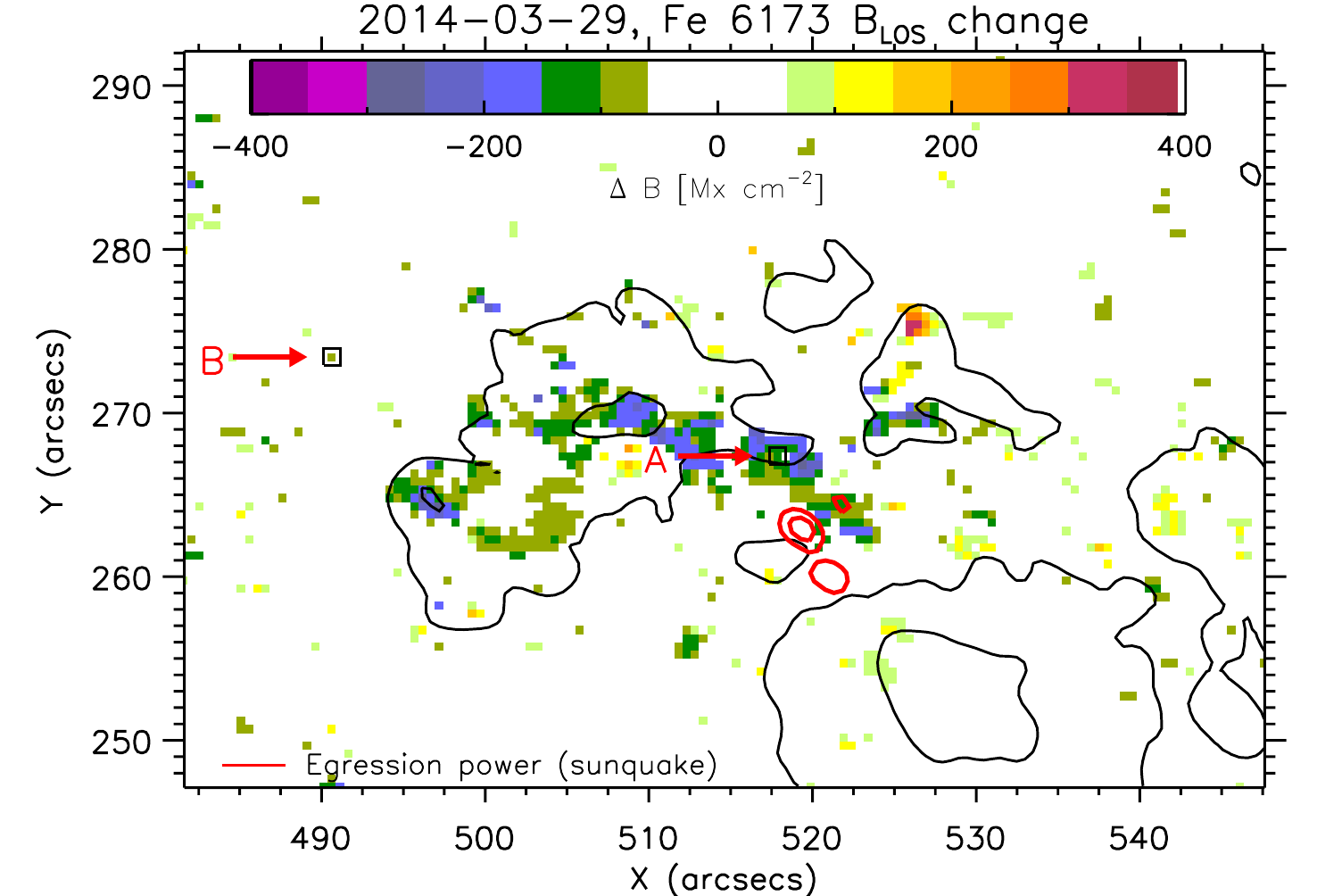}
     \includegraphics[width=.48\textwidth]{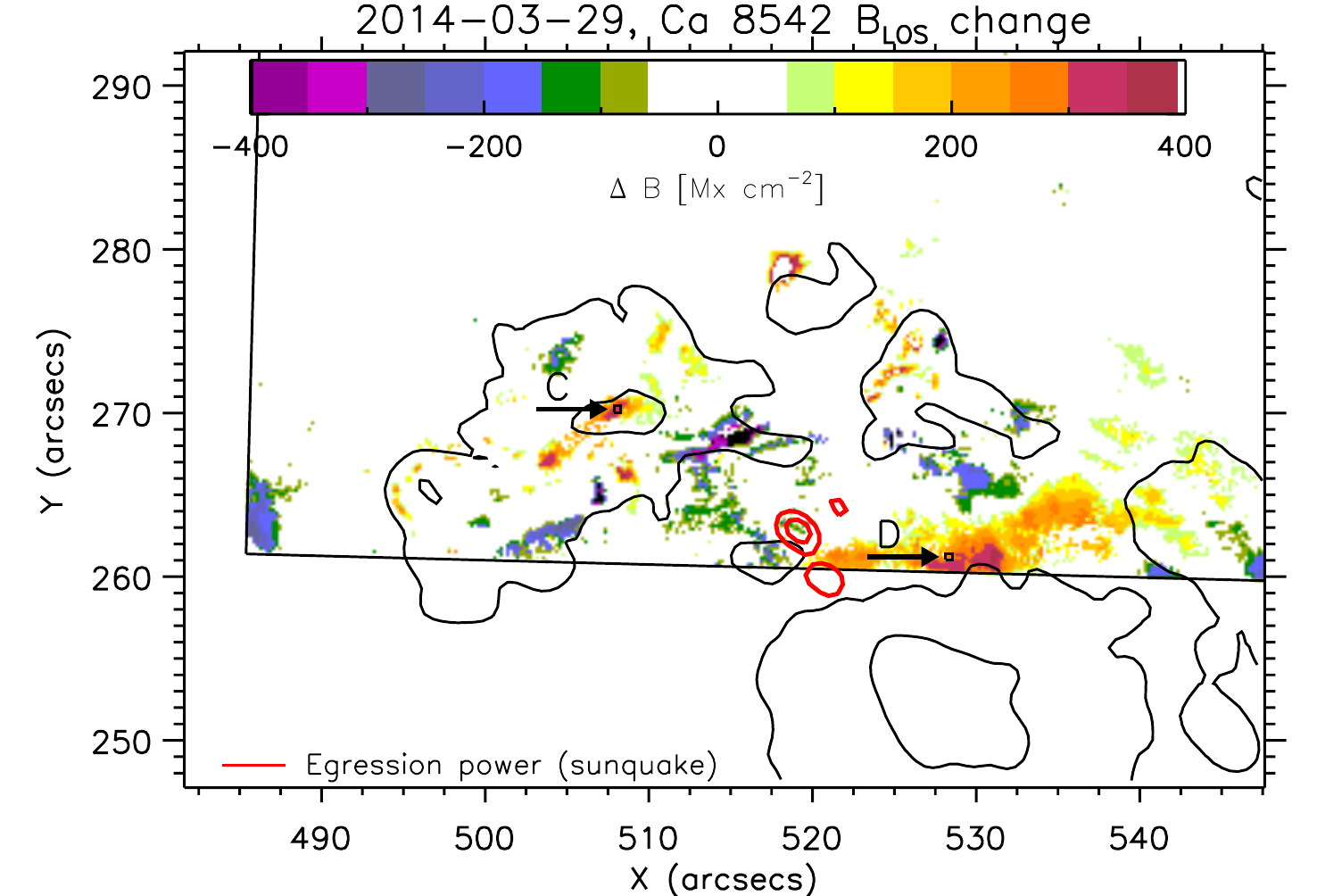}
   \includegraphics[width=.48\textwidth]{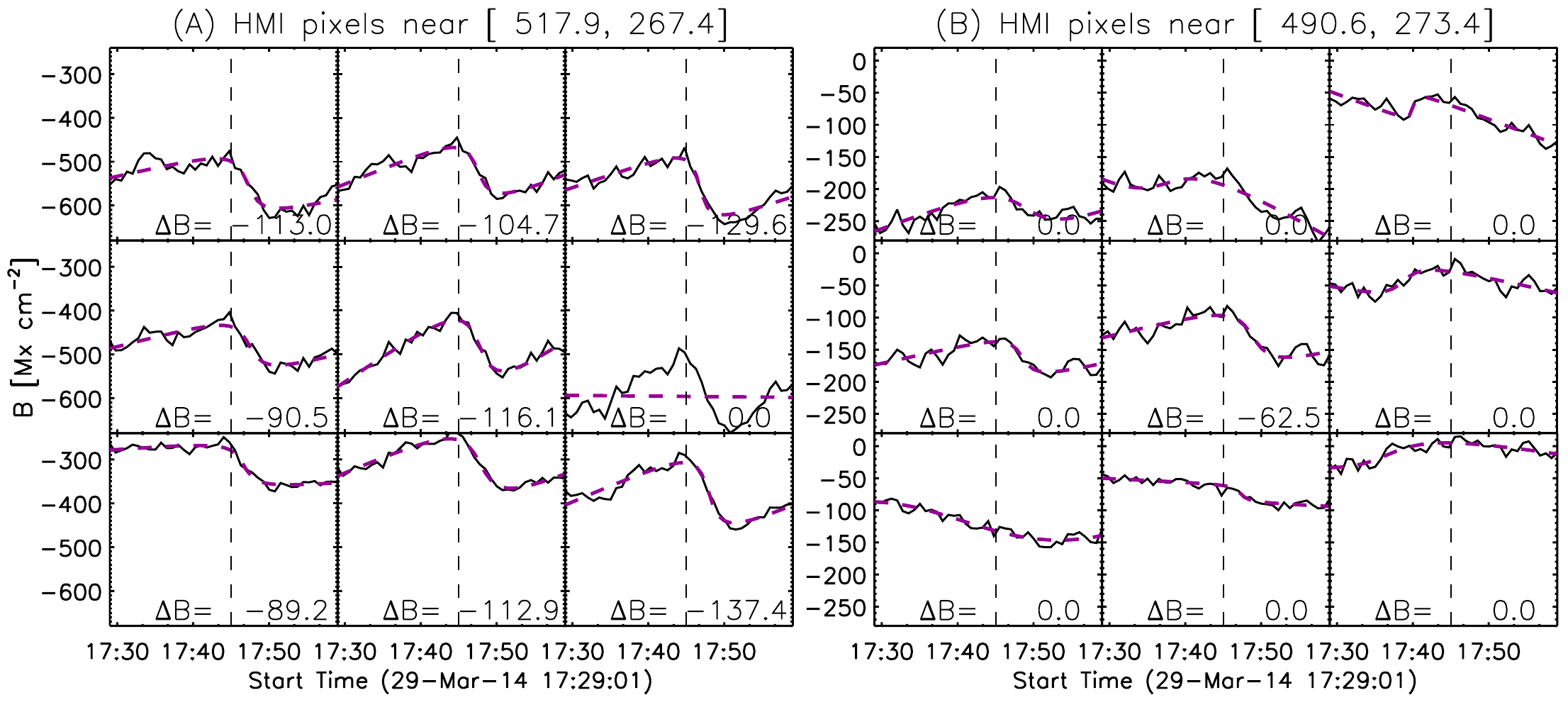}
     \includegraphics[width=.48\textwidth]{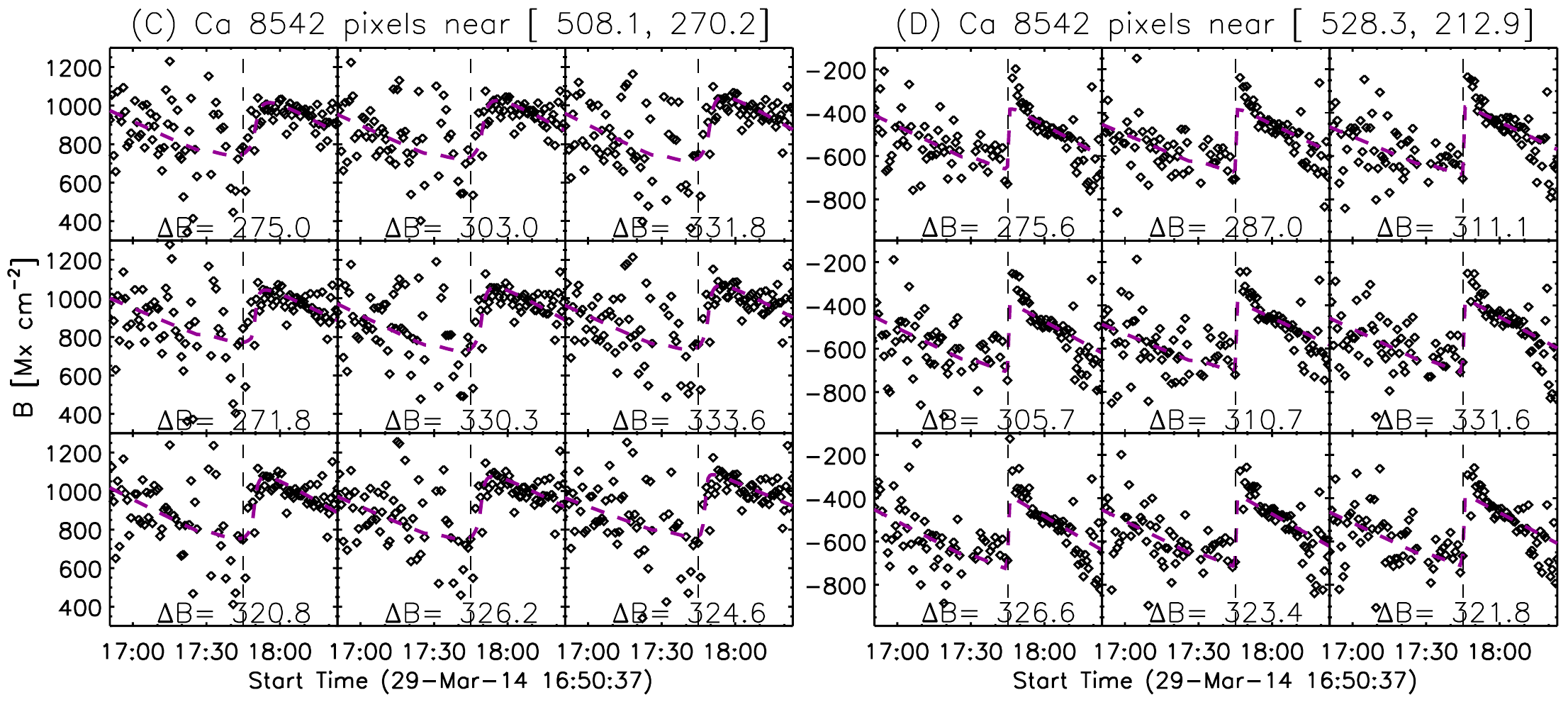}
   \caption{Top: Color-coded photospheric (left) and chromospheric (right) magnetic field changes drawn over contours of the sunspots/pores. Red contours represent the egression power of the (weak) sunquake \citep[see][]{judgeetal2014}. Arrows (marked with A-D) point to selected regions of 3x3 pixels whose profiles are shown in the bottom panels. Fits from Eq.~\ref{eq1} are shown in purple dashed lines. $\Delta$B denotes the value of the stepwise change. 
   Photosphere: Panels (A) shows clear stepwise changes of B$_{\rm LOS}$(t) and panels (B) show pixels in the quiet Sun. The middle pixel of (B) may appear like noise in the image, but it shows a stepwise change of -63 Mx cm$^{-2}$ at flare time while its neighboring pixels do not. Chromosphere (panels C and D): Clear stepwise changes of B$_{\rm LOS}$(t) are visible in both footpoints (evidence that they are footpoints is shown in e.g.~Fig.~\ref{inclination}).}
        \label{hminoise}
  \end{figure*}

  \begin{figure*}[tb] 
  \centering 
   \includegraphics[width=.48\textwidth]{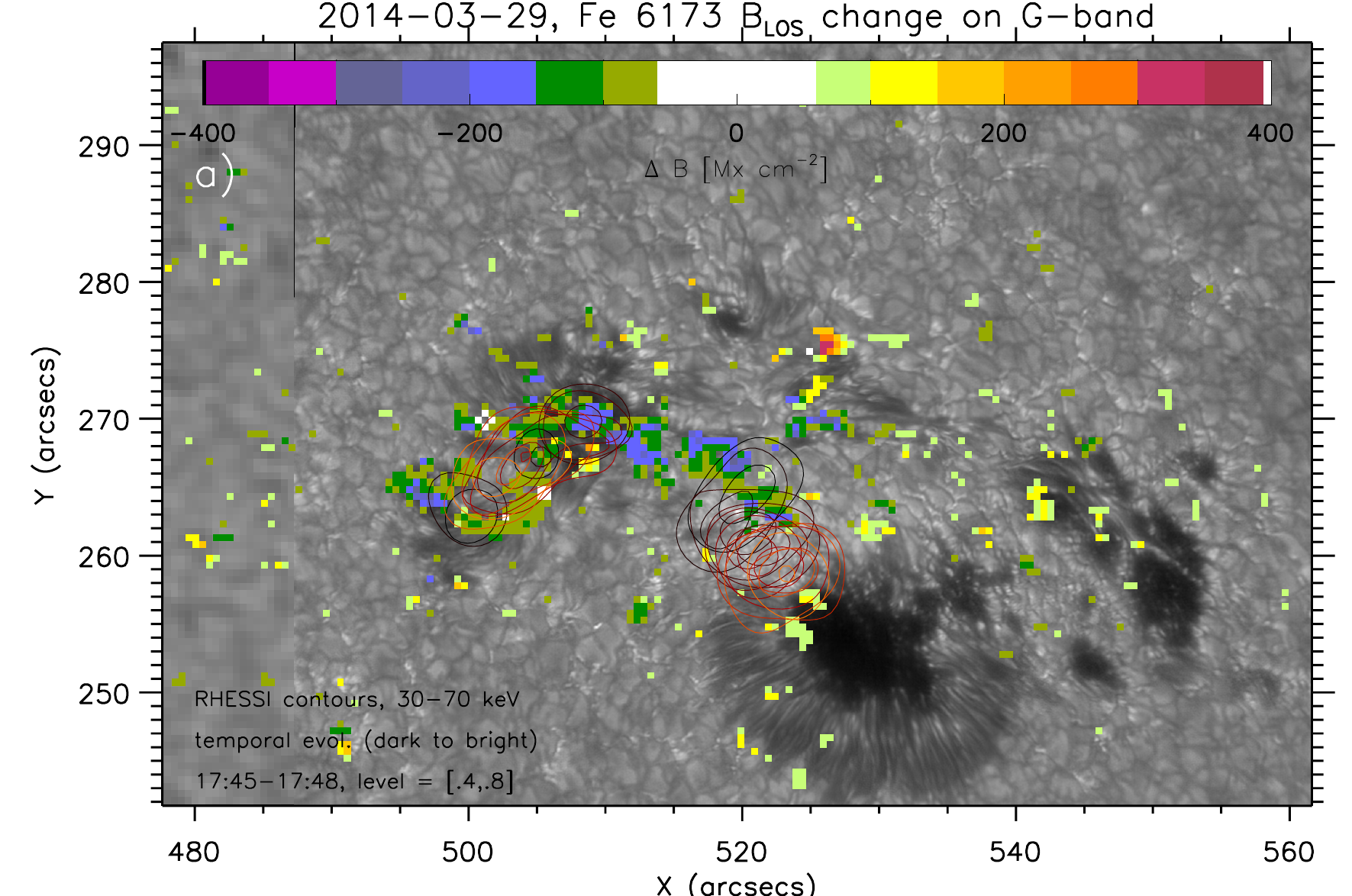}
   \includegraphics[width=.48\textwidth]{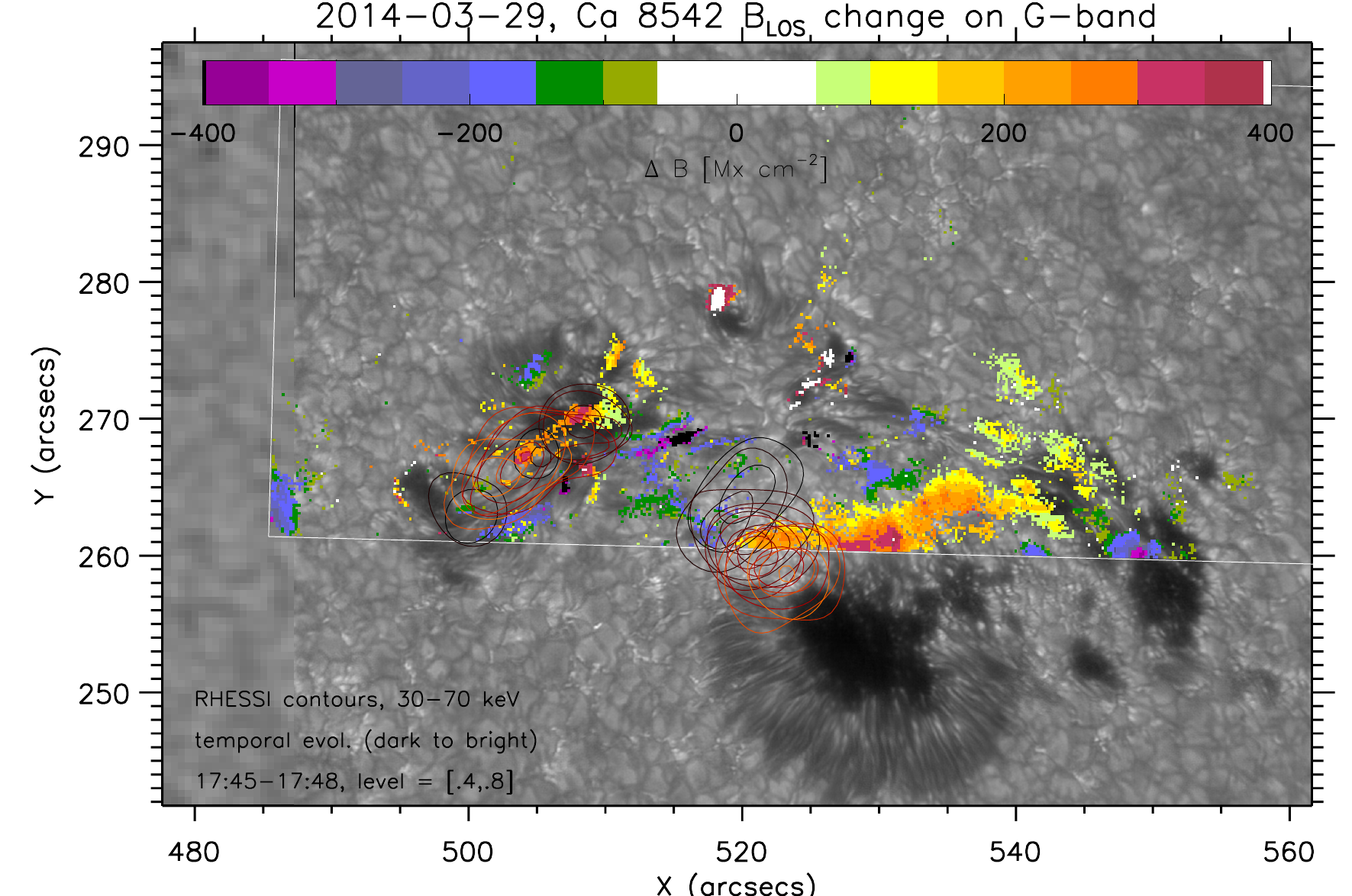}
   \includegraphics[width=.48\textwidth]{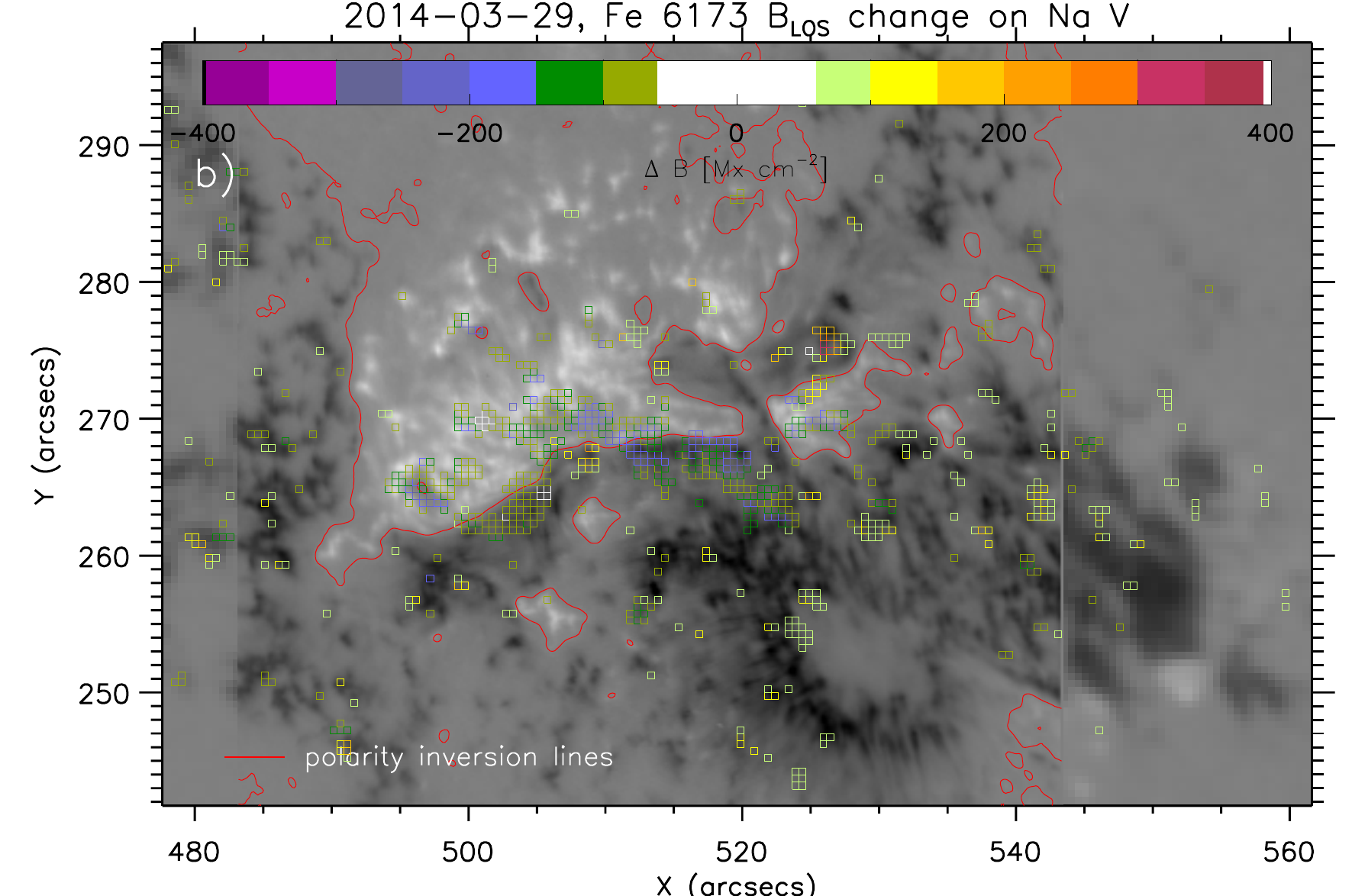}
   \includegraphics[width=.48\textwidth]{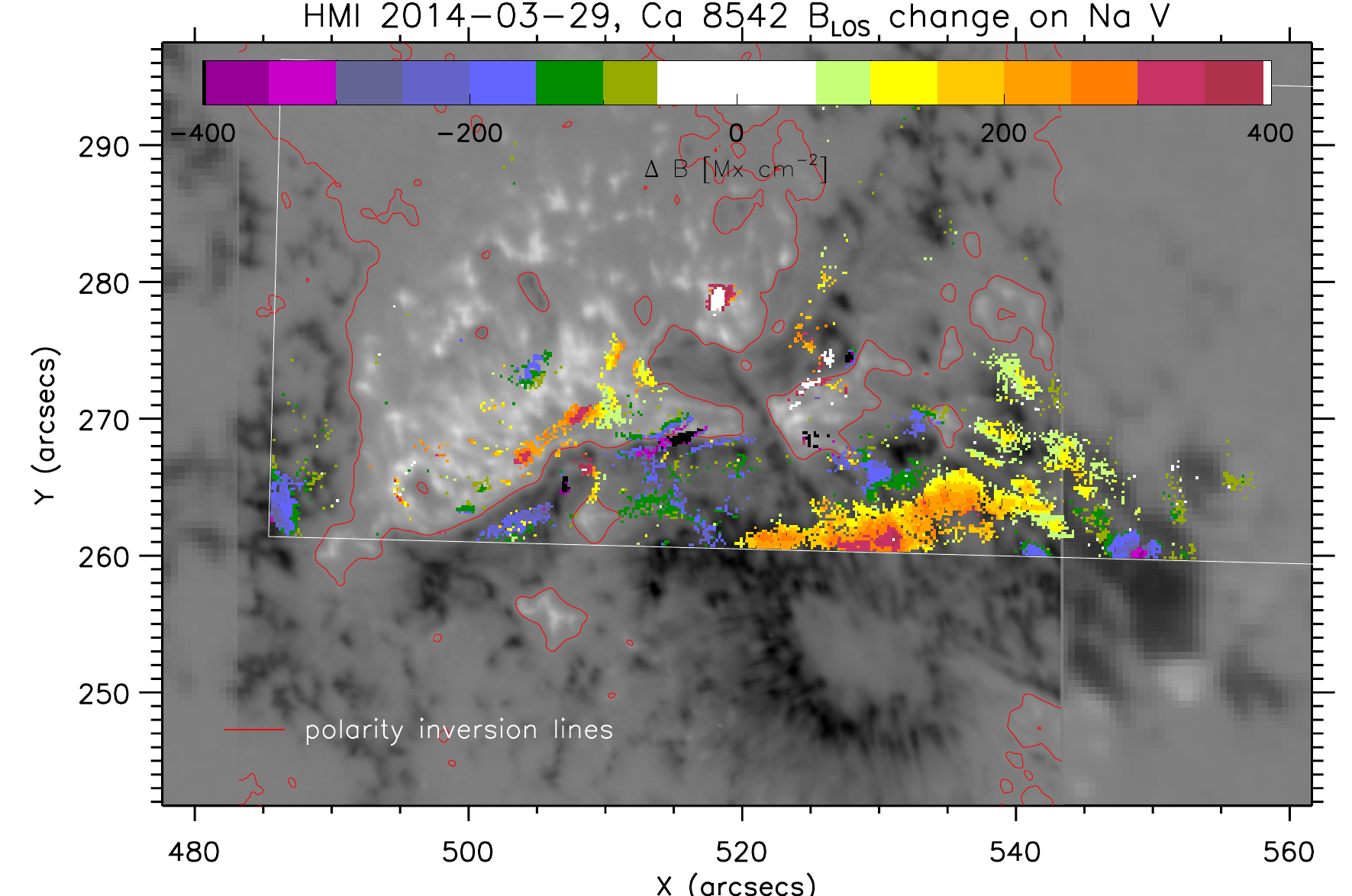}   
   \includegraphics[width=.48\textwidth]{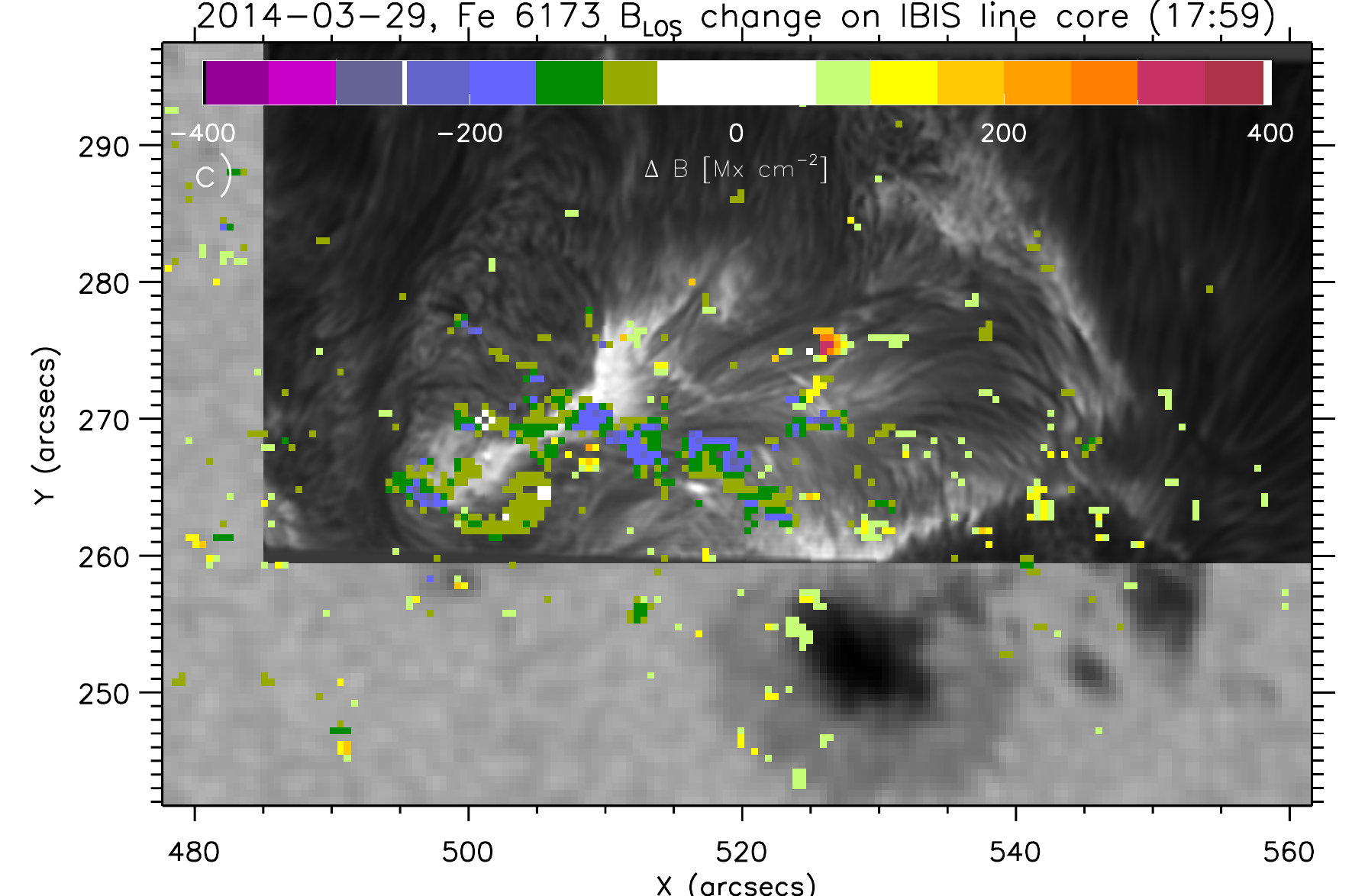}
   \includegraphics[width=.48\textwidth]{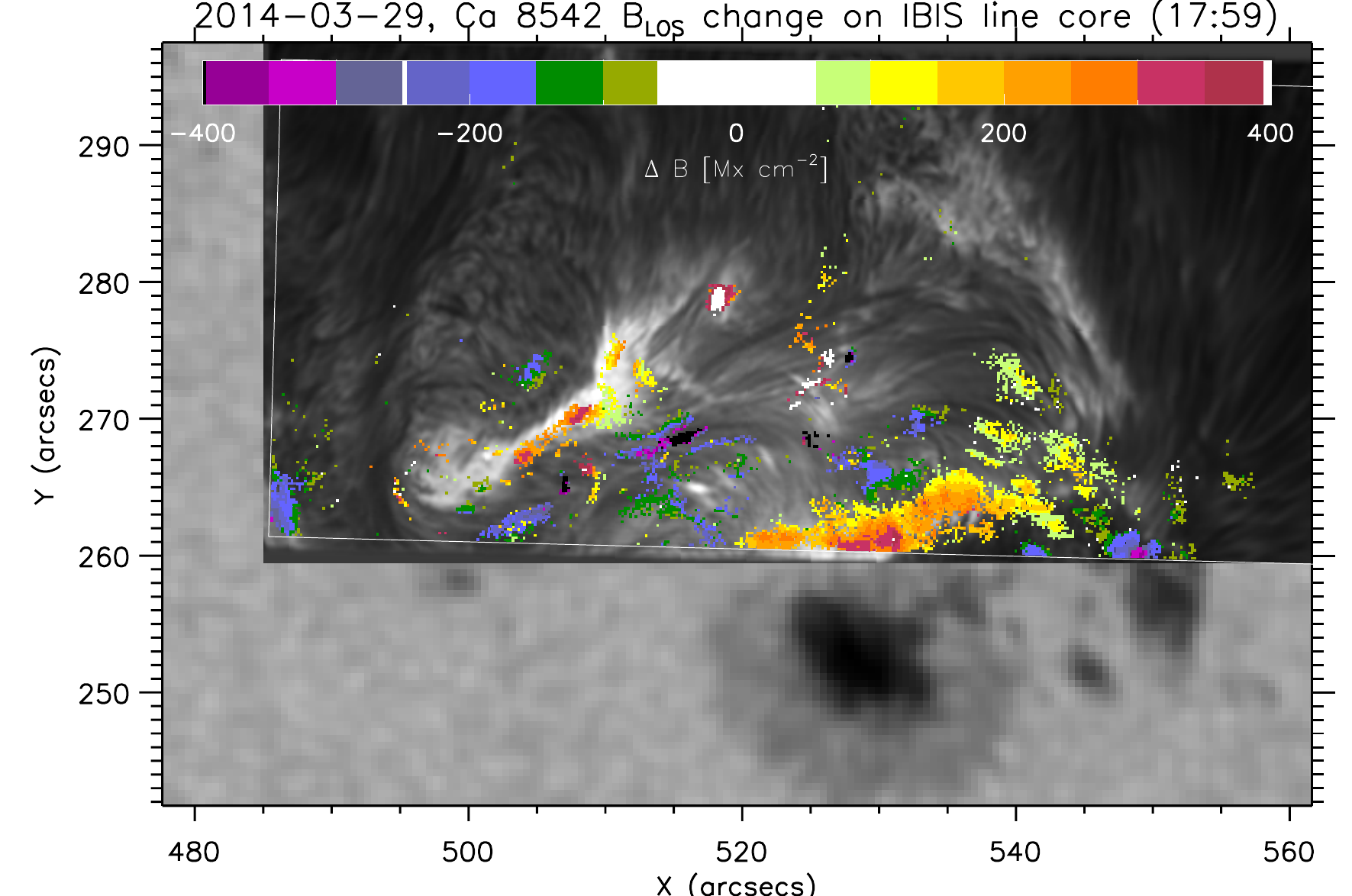}
    \caption{Left column: Photospheric changes of B$_{\rm LOS}$ derived from HMI with their size color-coded according to the colorbar. Right column: Chromospheric changes of B$_{\rm LOS}$ derived from IBIS (color scale clipped to match photospheric $\Delta$B). a) The background image shows the photospheric intensity from HMI and a DST/G-band overlay for higher spatial resolution. IBIS' FOV is indicated with white lines. The contours (40\%,80\%) correspond to the temporal evolution (dark to bright) of the 30-70 keV HXR emission from RHESSI. b) background: HMI magnetogram with an overlay of Hinode's \ion{Na}{1} Stokes $V$ for higher resolution. Red lines indicate polarity inversion lines. c) background: speckle-reconstructed line core image of \ion{Ca}{2} 8542.1 \AA\ showing that chromospheric changes occur mainly co-spatial to the ribbons.}
        \label{loc1}
  \end{figure*}

 \begin{figure*}[!ptbh] 
  \centering 
   \includegraphics[width=.47\textwidth]{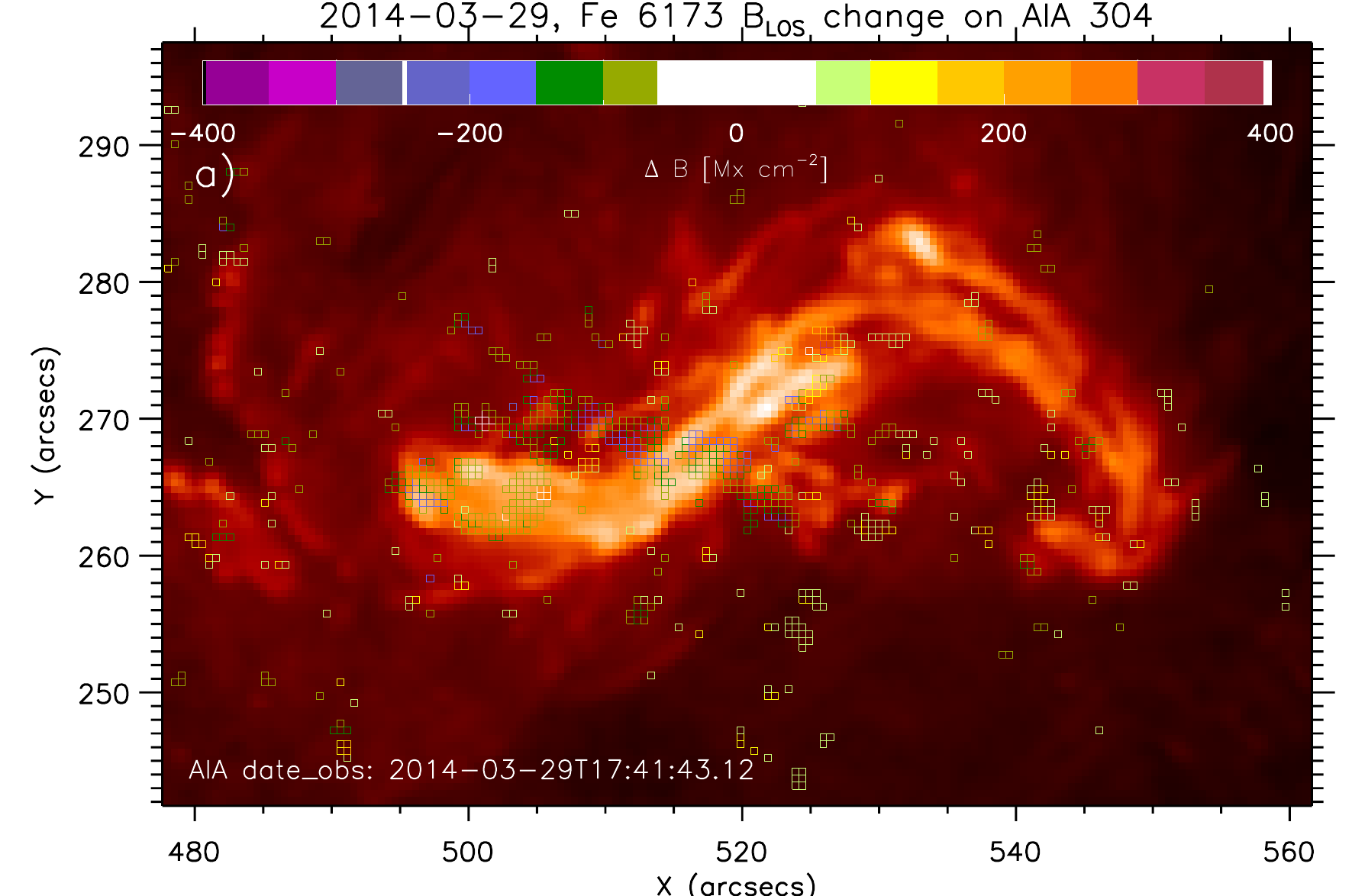}
   \includegraphics[width=.47\textwidth]{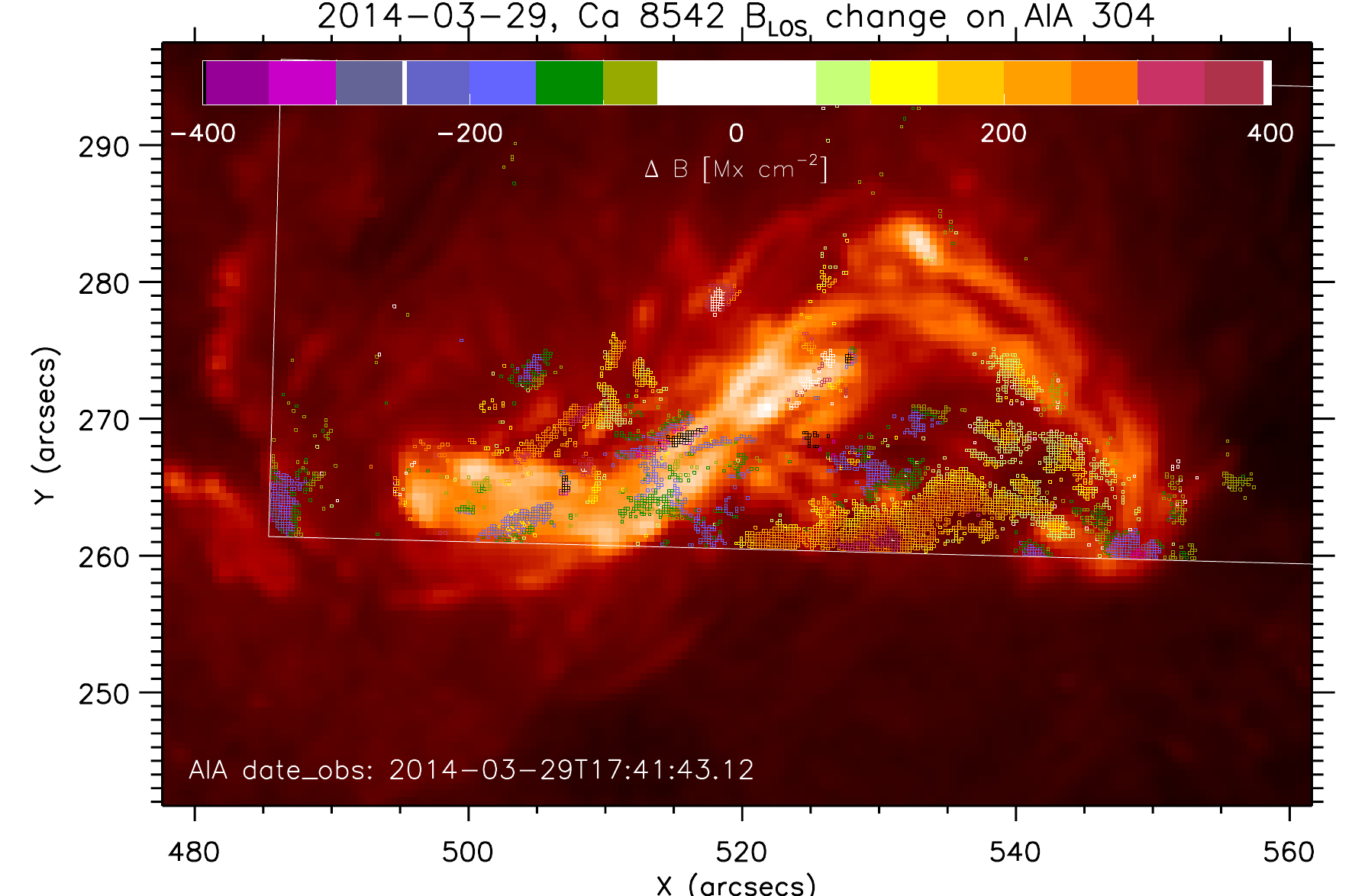}
   \includegraphics[width=.47\textwidth]{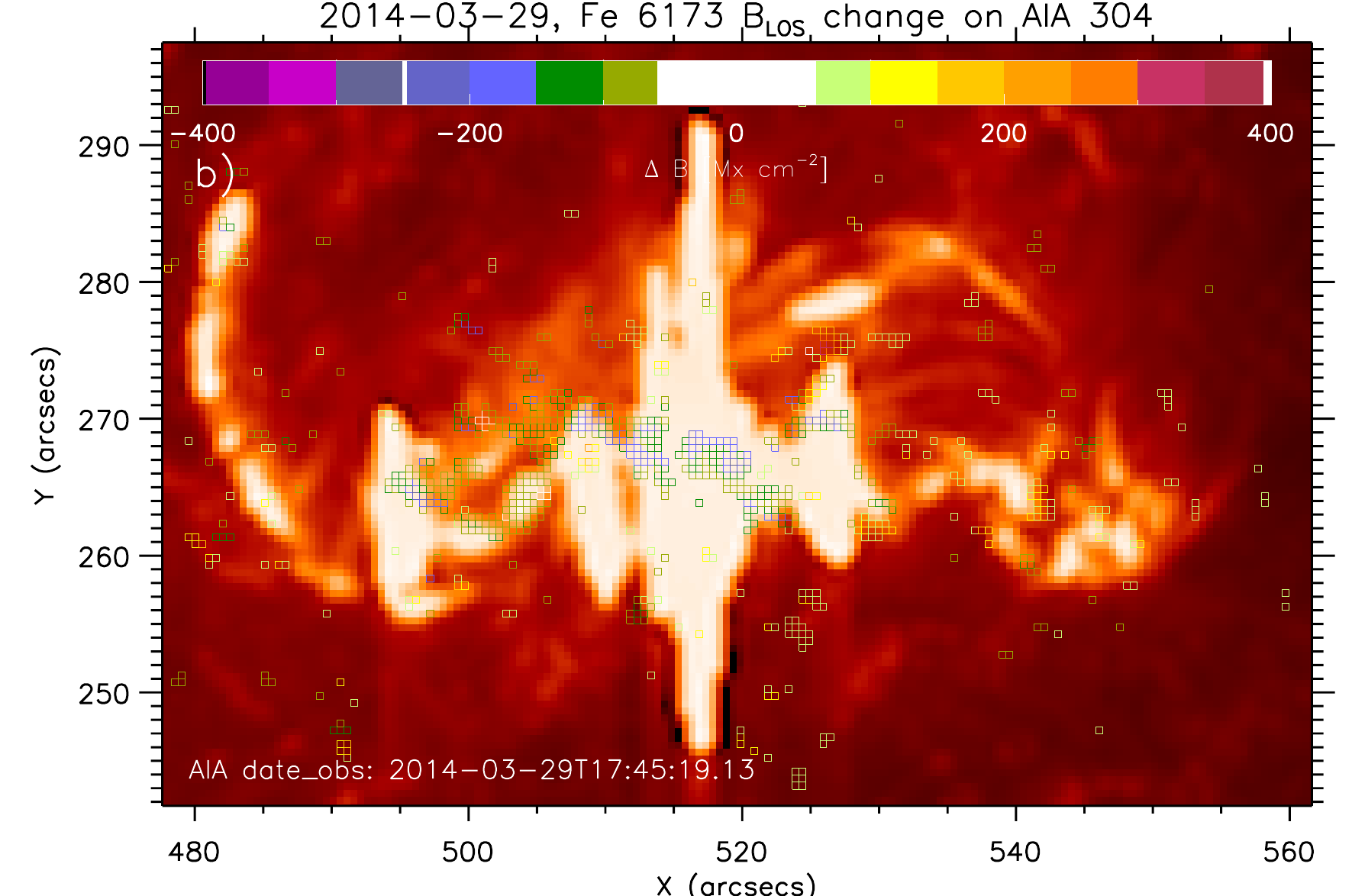}
   \includegraphics[width=.47\textwidth]{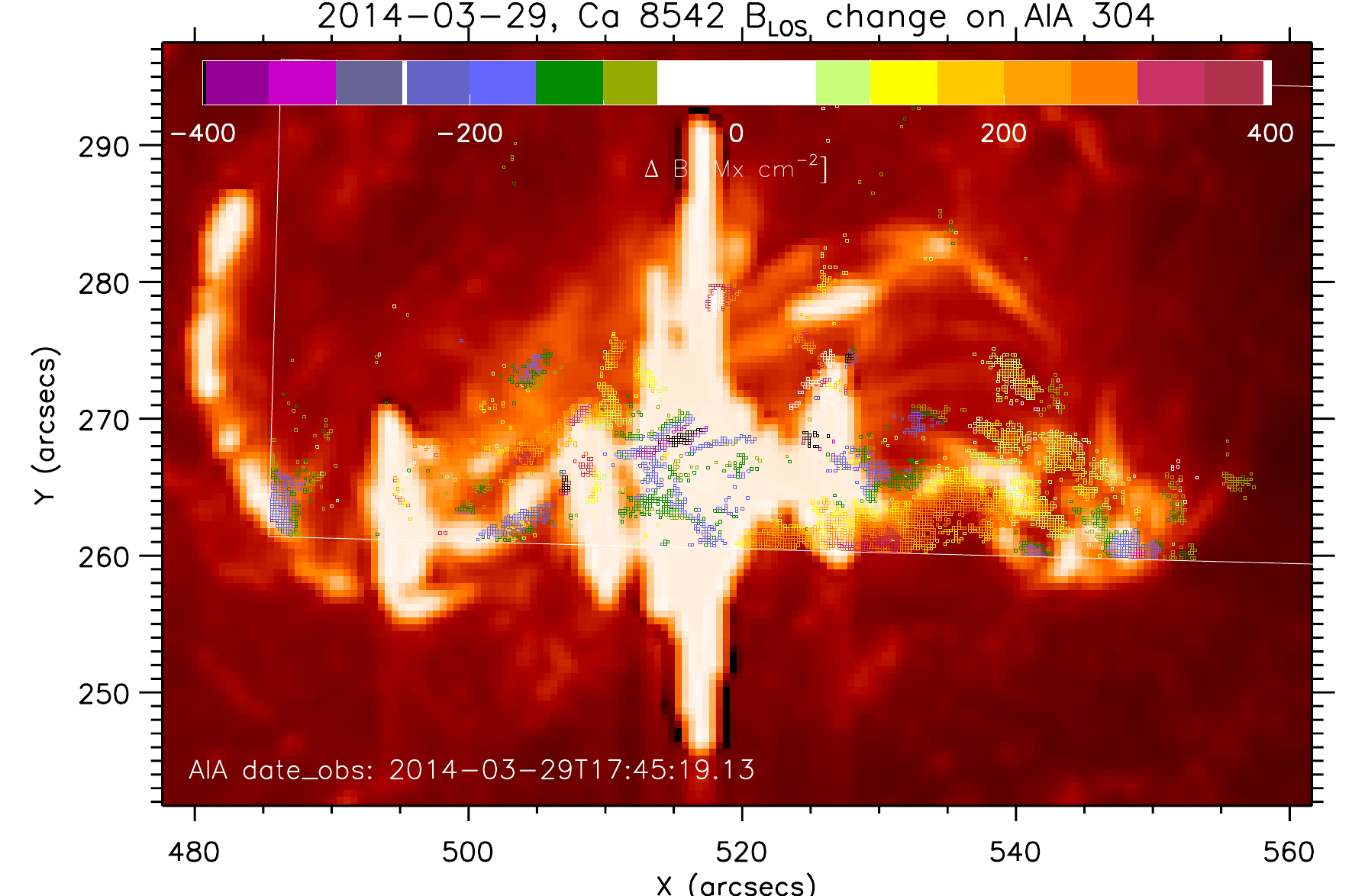}
    \includegraphics[width=.47\textwidth]{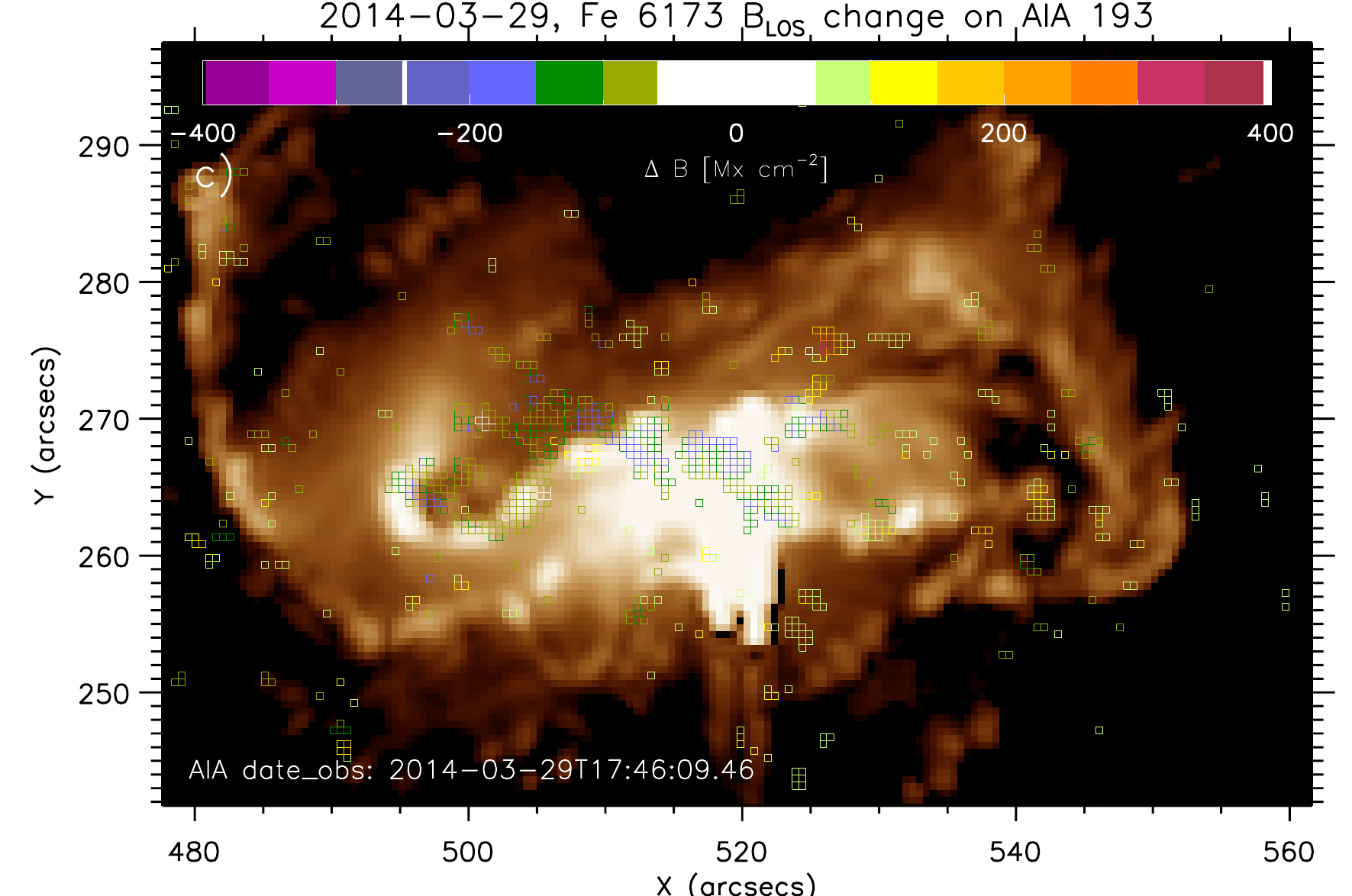}
   \includegraphics[width=.47\textwidth]{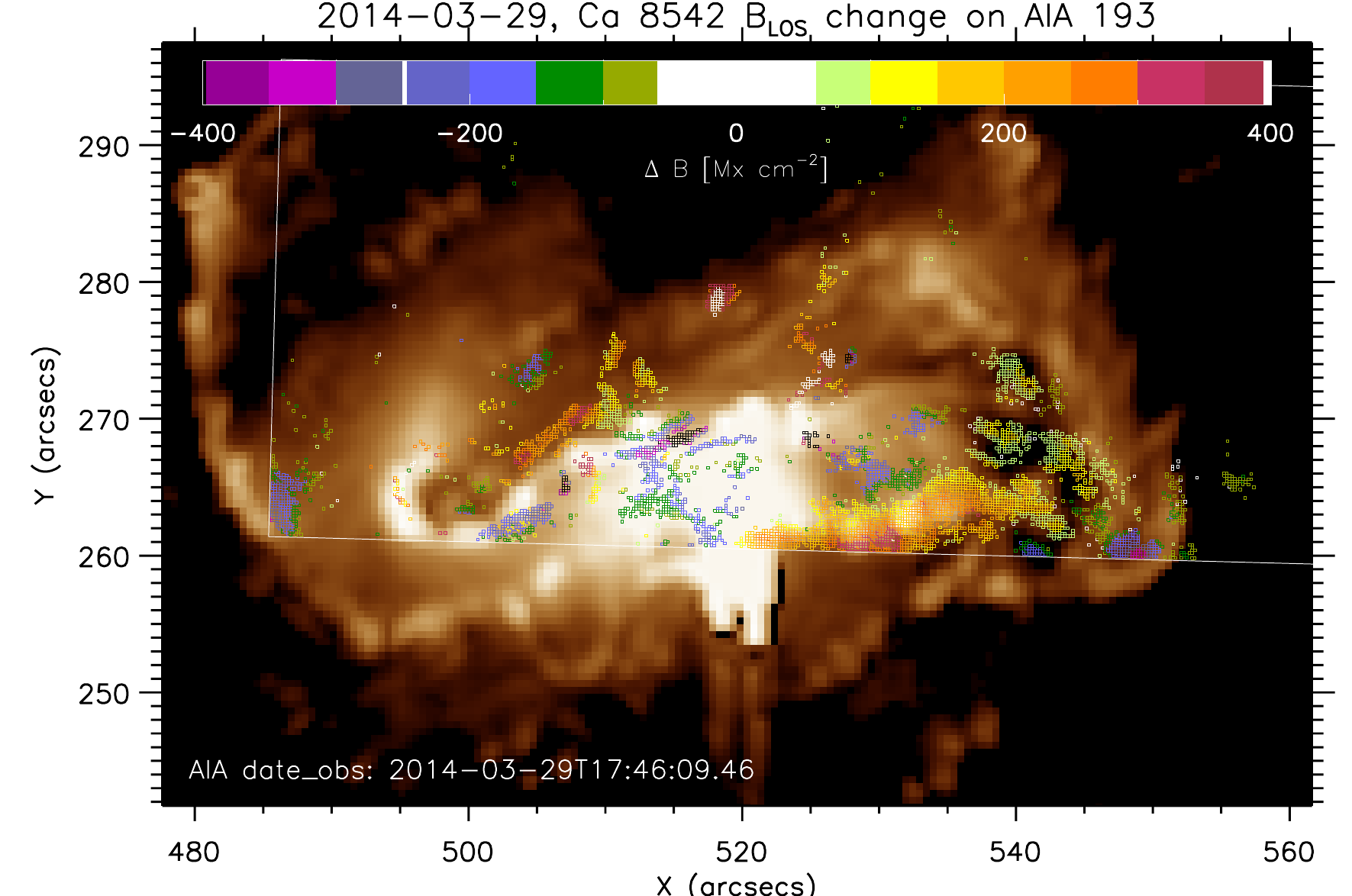}
   \includegraphics[width=.47\textwidth]{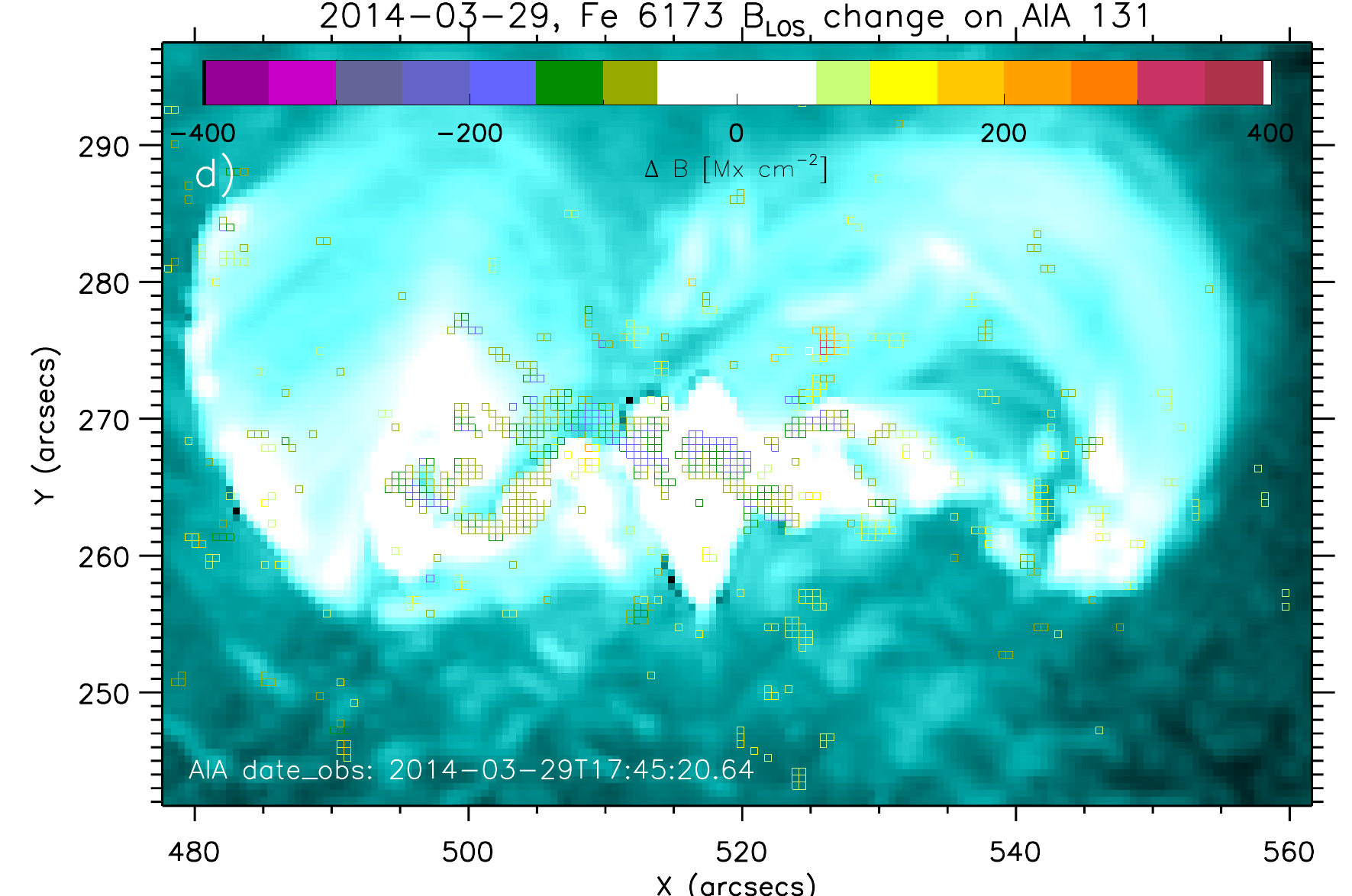}   
   \includegraphics[width=.47\textwidth]{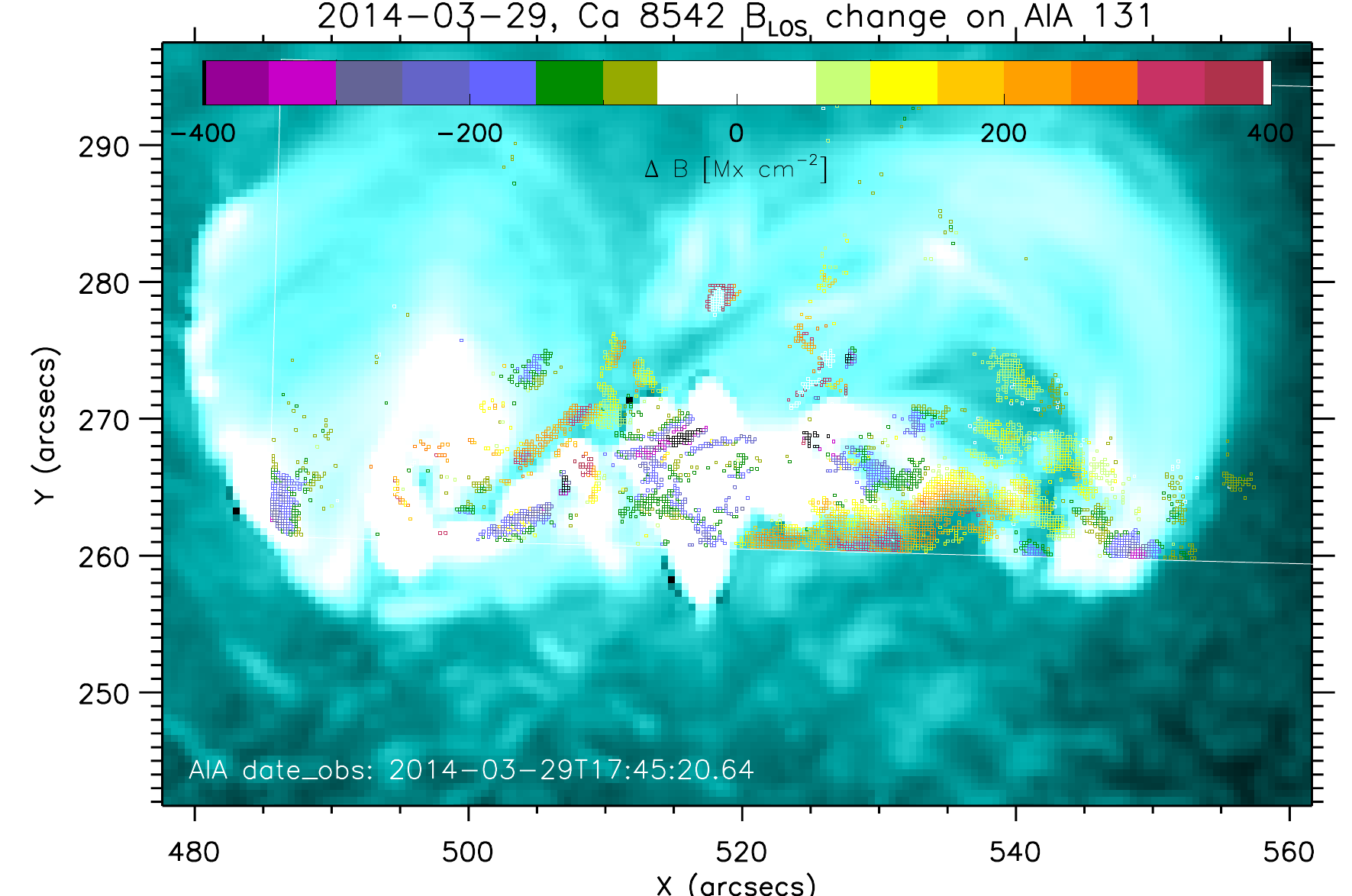}
  \caption{Same as Fig.~\ref{loc1}: photospheric (left) and chromospheric changes (right). a) The background image shows AIA 304 at 17:41:43, just when the filament was erupting (bright structure). Some changes of B$_{\rm LOS}$ occurred later near its footpoints. b) background: AIA 304 at 17:45:19 UT. c) background: AIA 193 at 17:46:09~UT. d) background: AIA 131 at 17:45:20 UT showing the coronal loop structure.}
        \label{loc2}
  \end{figure*}

\section{Results}

\subsection{Comparison of $\Delta$B$_{\rm LOS}$}
The retrieved values of the observed step size ($\Delta$B$_{\rm LOS}$) are shown as histograms in Fig.~\ref{histogram} for the photosphere (left) and the chromosphere (right). Values below $\pm$~60~Mx cm$^{-2}$ are not shown (dashed region) because of low S/N. It can be seen clearly that the chromosphere shows  larger changes. The numbers of changes are not directly comparable because of the different spatial resolution and FOVs of the instruments. There is an asymmetry in the chromospheric changes with more positive changes. They mostly arise from an area near the negative polarity sunspot, where $\left| {\rm B}_{\rm LOS}\right|$ decreased after the flare. The maximum observed photospheric changes are $\sim$320 Mx cm$^{-2}$, those of the chromosphere $\sim$640 Mx cm$^{-2}$. The apparent decrease below $\pm \approx$100 Mx cm$^{-2}$ in the chromospheric histogram arises from the fact that many temporal sequences of B(t) are noisy with a standard deviation comparable to the step size. Such pixels were excluded from the analysis and it is assumed that many stepwise changes below 100 Mx cm$^{-2}$ are invisible in our chromospheric data.

\subsection{Locations of changes of B$_{\rm LOS}$}

The locations of the changes are shown in Fig.~\ref{hminoise}, \ref{loc1}, and \ref{loc2}. Figure~\ref{hminoise} shows the photospheric (left) and chromospheric (right) changes, including example plots of B$_{\rm LOS}$(t) for two selected regions (3x3 pixels, marked by arrows). Photospheric region A (pixels near [517.9,267.4]\arcsec) shows clear stepwise changes exactly at the start of the flare (denoted by the vertical dashed lines). $\Delta$B denotes the value of the magnetic field change, a zero means that the pixel did not show a stepwise change or that the fit failed. Even in the quiet Sun (e.g. pixels near [490.6,273.4]\arcsec, panels B), there are photospheric stepwise changes visible right at flare time. While they may look like noise in the image, they were not excluded because they passed all criteria for stepwise changes. The chromospheric examples in panels C and D show clear stepwise changes in both footpoints of the flare. Also drawn is the location of the weak sunquake (egression power; red contours) according to \citet{judgeetal2014}. It can be seen that the main source is near, but does not coincide with the magnetic field changes neither in the photosphere, nor in the chromosphere. A very weak potential source at [522, 265]\arcsec (it is unclear if this source is an actual sunquake because it is so weak; A. Donea, private communication) may coincide with some of the weaker photospheric changes, but the strongest magnetic field changes definitely do not show a corresponding sunquake.

Colored pixels in Figures~\ref{loc1} and \ref{loc2} show the size and location of changes of B$_{\rm LOS}$, in the left column for the photosphere and in the right column for the chromosphere. Each row has a different background image with its origin given in the titles. 

In the photosphere, changes of B$_{\rm LOS}$ are concentrated along the neutral lines (red contours, Fig.~\ref{loc1}b) and near the pores. Many of them and especially the strongest ones do not correspond to locations of the HXR contours whose temporal evolution from 17:45-17:49 is plotted in Fig.~\ref{loc1}a. In fact many changes seem to occur right \textit{next} to the HXR emission, especially in the south-western footpoint, which also had the stronger HXR emission. Only the very early HXR emission (black contours in Fig.~\ref{loc1}a, corresponding to 17:45:25-17:45:45) is co-spatial to some photospheric changes, mostly in the 40\% HXR contours of the north-eastern footpoint. That the early HXR emission is closer to the magnetic field changes has also been noticed by \citet{burtsevaetal2015} for other flares. Photospheric changes do seem to occur in places with strong emission in AIA, especially the circularly-shaped eastern part of the changes has a correspondence in e.g.\,AIA 304 (Fig.~\ref{loc2}b). The reverse is not true, strong emission in AIA does not necessarily indicate locations of photospheric magnetic field changes. 61\% of photospheric changes \textit{decrease} the absolute value of B$_{\rm LOS}$.

The chromospheric changes of B$_{\rm LOS}$ encompass a larger area than the photospheric changes. While those in the northern sunspot and near the pores are somewhat co-spatial, there is a large area near the southern sunspot that shows significant and large changes only in the chromosphere. This area corresponds to the southern flare ribbon, seen in the Ca 8542 \AA\ line core image in Fig.~\ref{loc1}c. That panel also shows brightenings at loop tops (e.g. [517, 265]\arcsec or [526, 270]\arcsec) and these loops seem to connect both ribbons where large chromospheric changes were observed. The area is on the northern side of the large sunspot (cf. Fig.~\ref{loc1}a) and did not show any regular penumbra. Throughout the observations, small filamentary structures and granules kept emerging. 

Flare-related changes generally seem coherent in area. Single pixel changes (e.g.~edge of umbra/penumbra or penumbra/quiet Sun) could also be due to regular solar evolution. The chromospheric changes clearly show that there is no correlation to the location of HXR emission. This was also tested by plotting the integrated RHESSI HXR and GOES lightcurves and overplotting the number of changes at each time step ($t_0$ in Eq.~\ref{eq1}), where also no correlation was visible.
The majority of chromospheric B$_{\rm LOS}$ changes also does not lie on a neutral line. Nearly all locations with chromospheric changes exhibit enhanced emission in AIA at some point during the flare (c.f.~Fig.~\ref{loc2}d for AIA 131 emission near the large area of changes). It seems that chromospheric changes predominantly occur near footpoints of coronal loops and near the western footpoint of the erupting filament (c.f.~blue area in Fig.~\ref{loc2}a near [550,260]\arcsec), which seems reasonable considering that these loops visibly change during the flare. Similarly to photospheric changes, 62\% of chromospheric changes \textit{decrease} the absolute value of B$_{\rm LOS}$. The whole south-western patch shows a decrease, while a large fraction of the north-eastern changes show increases of B$_{\rm LOS}$ (not shown in figures, can be deduced from Fig.~\ref{loc1}b).

  \begin{figure*}[!tbh] 
  \centering 
   \includegraphics[width=.87\textwidth]{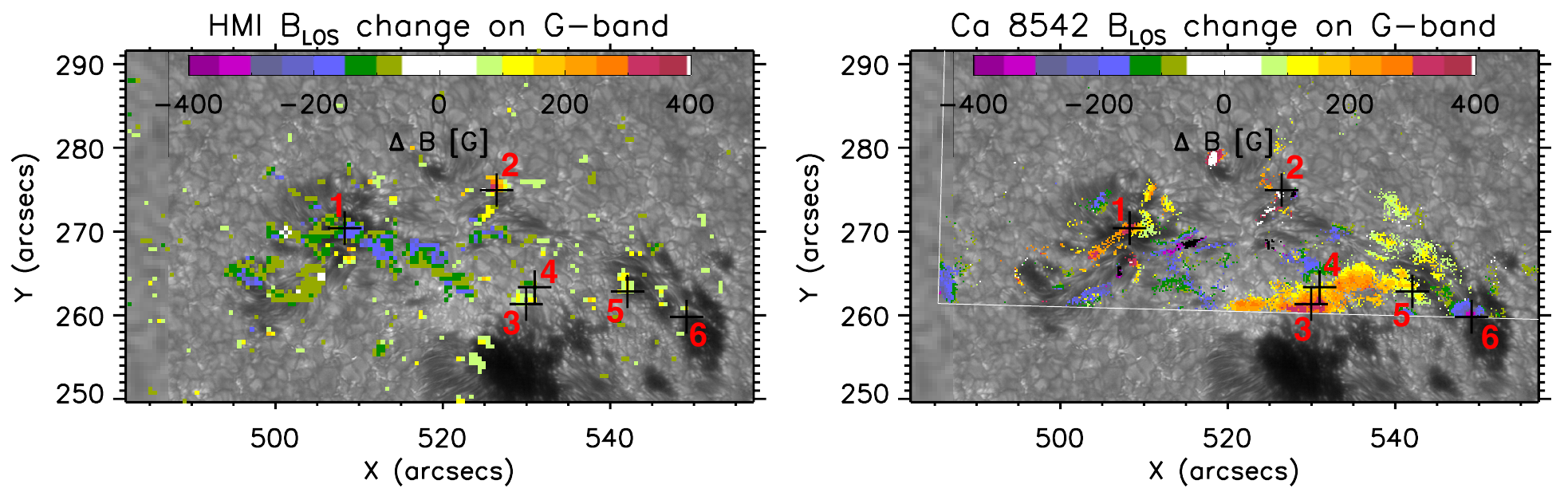}
  { \includegraphics[width=.98\textwidth]{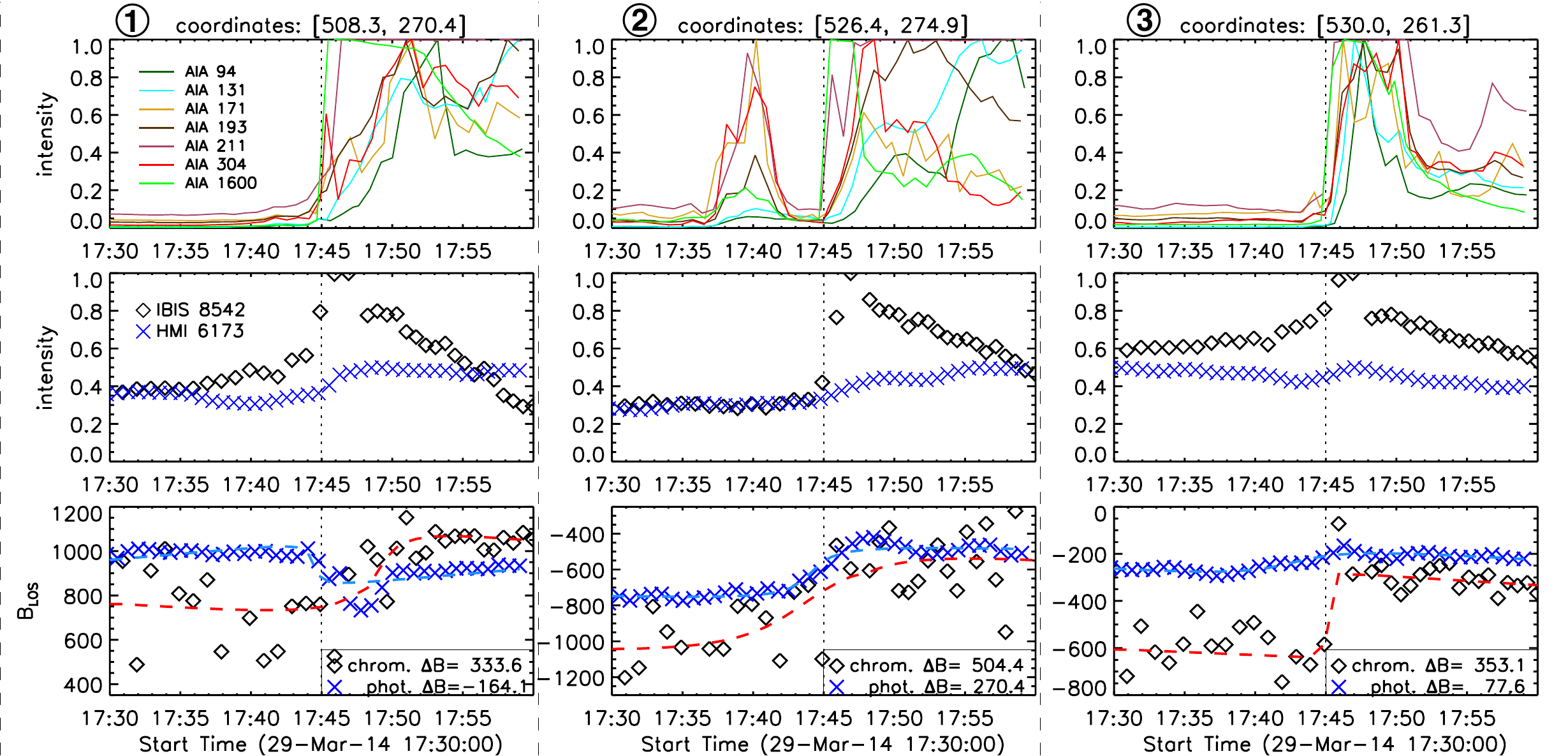}}
  { \includegraphics[width=.98\textwidth]{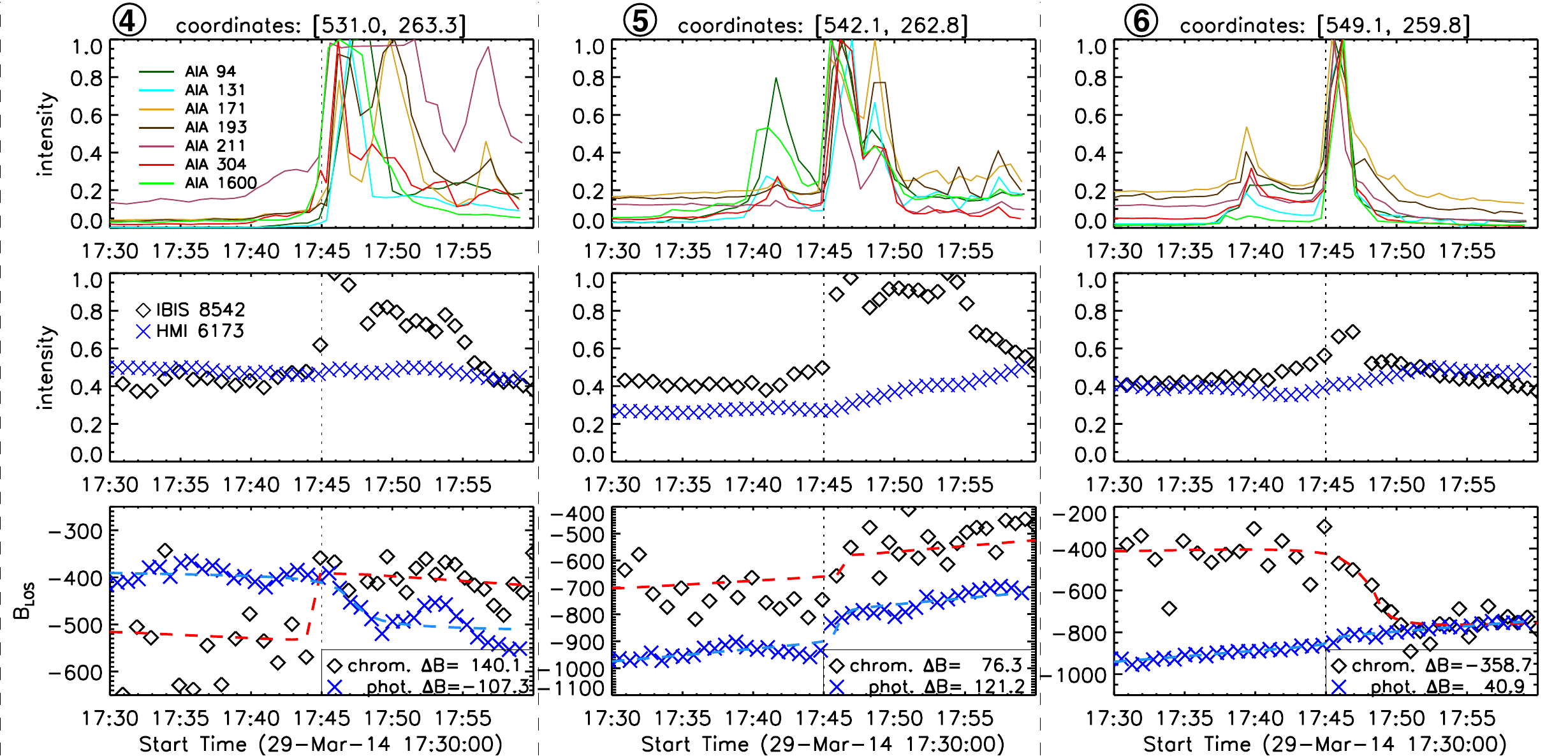}}
   \caption{Examples of temporal evolutions of AIA, HMI and IBIS for six example pixels whose locations are indicated with crosses and numbers in the top images. The panels below the figures are organized into groups of 3 vertical panels with the coordinates/number of each selected pixel given in the title of each set. The top panels of each set show AIA intensities (arbitrary units), the middle panels HMI (blue crosses) and IBIS (black diamonds) intensities (arbitrary units) and the bottom panels their magnetic field evolution (again HMI in blue, IBIS in black). Dashed lines indicate stepwise fits according to Eq.~\ref{eq1}. Magnetic field changes in the photosphere and chromosphere behave seemingly independently, sometimes with opposite signs of jumps.}
        \label{lightcurves}
  \end{figure*}
  
  \begin{figure*}[tb] 
  \centering 
   \includegraphics[width=.45\textwidth]{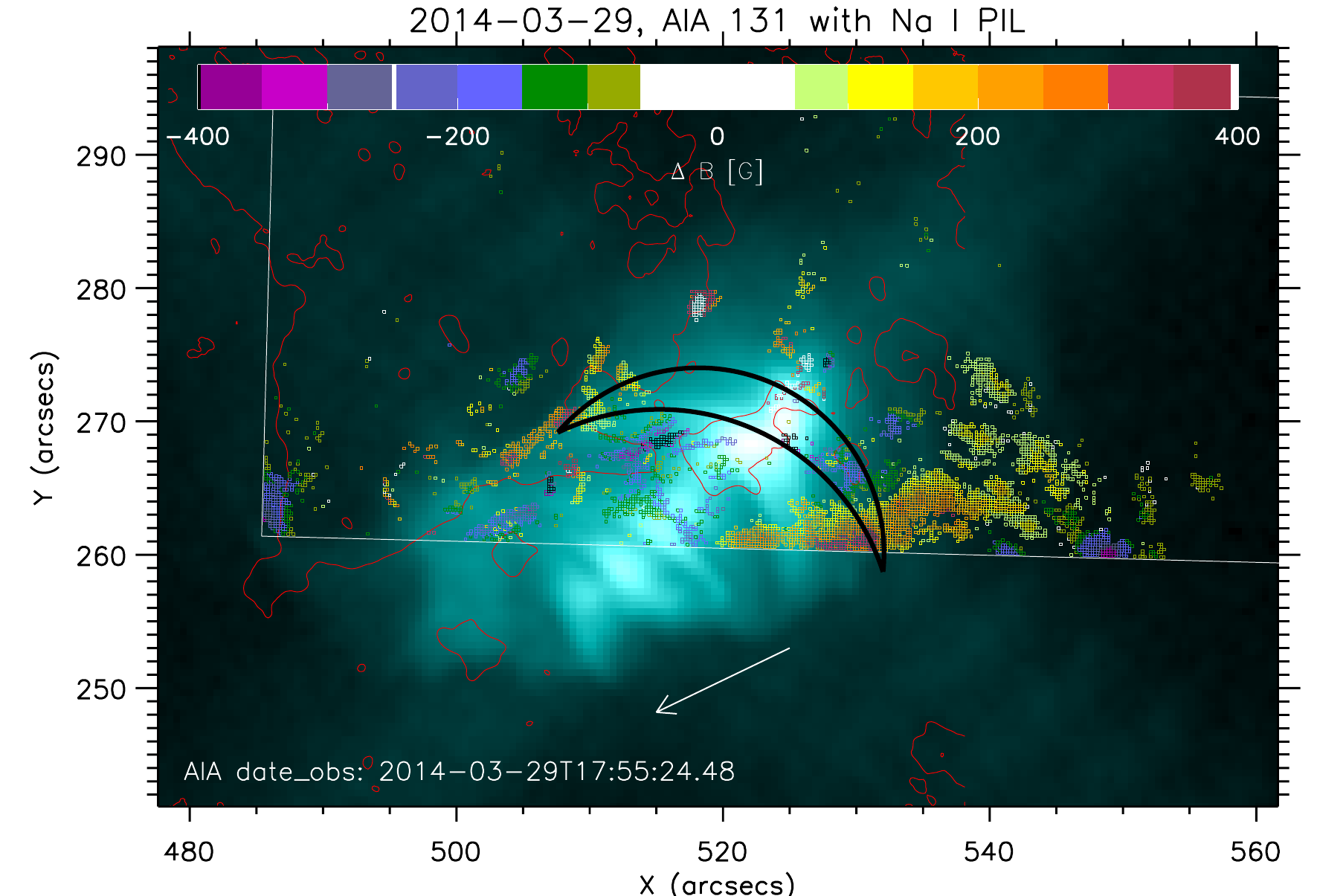}
   \includegraphics[width=.54\textwidth,bb=15 200 650 550,clip]{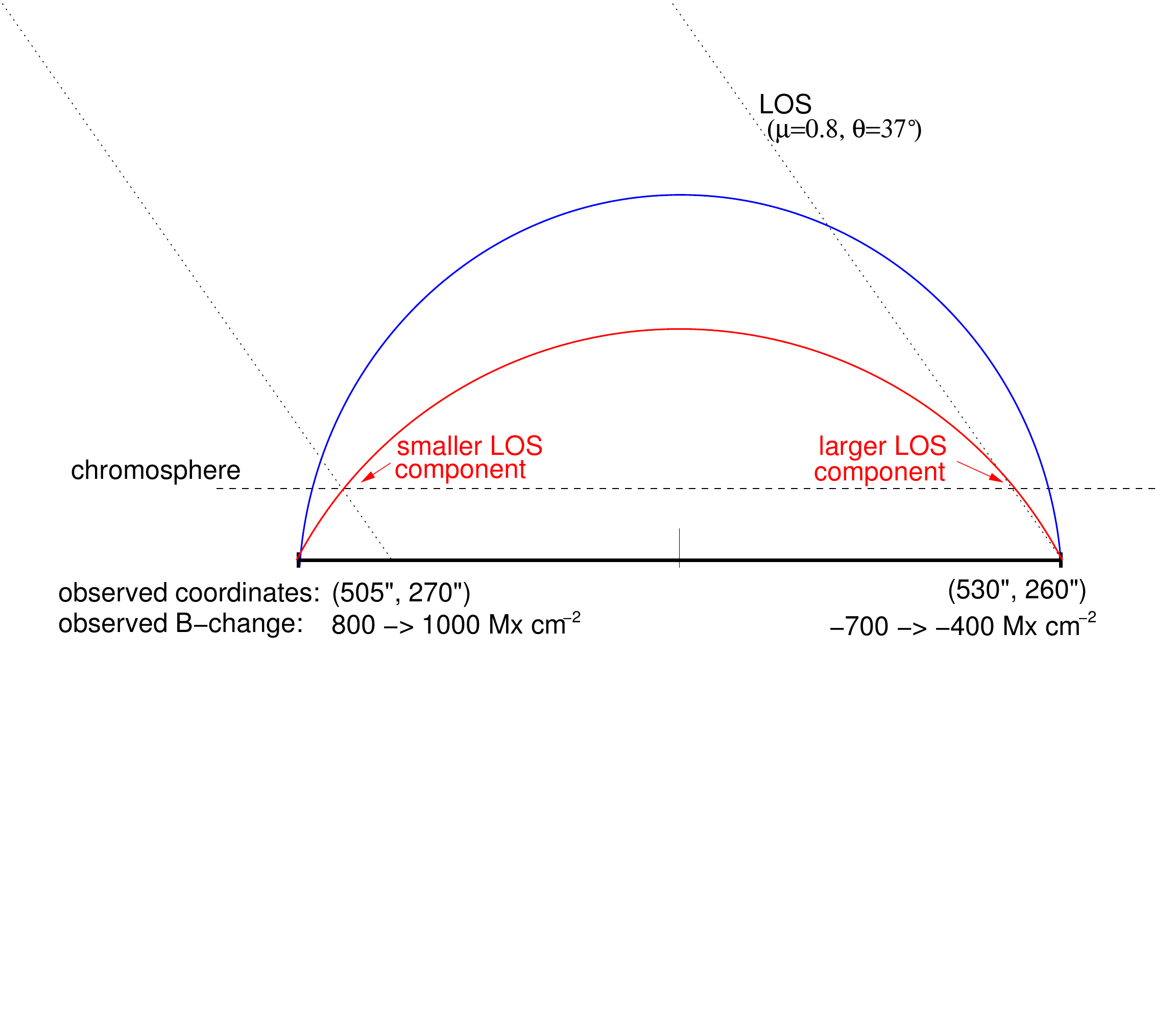}
   \caption{Left: AIA 131 image 10 min after flare maximum to show post-flare loops that trace the coronal magnetic field structure. Overplotted are the polarity inversion lines derived from Hinode (red), the chromospheric field changes (see color bar), the direction to disk center (white arrow) and two hand-drawn loops that represent potential connection sites and loop geometries. Right: Schematic drawing of the two marked loops. The LOS direction is indicated with dotted lines. The chromospheric height (dashed) was estimated from the observed loop base length ($\sim$25\arcsec). Compared to the larger blue loop, the left (eastern) side of the smaller red loop has a smaller LOS component of B. If the loop contracted, B$_{\rm LOS}$ should therefore decrease. However, the opposite was observed, as indicated by the values in the figure.}
        \label{inclination}
  \end{figure*}

\subsection{Timing of changes of B$_{\rm LOS}$}
To investigate the timing of the changes B$_{\rm LOS}$(t), the white-light emission, and emission in AIA, we plot lightcurves for different pixels in Fig.~\ref{lightcurves}. The top panels indicate the selected six pixels with plus signs. The light curves are grouped in sets of three vertically arranged panels with the top panels showing AIA, the middle panels the HMI continuum intensity (``white light'', blue crosses) and \ion{Ca}{2} 8542.1 \AA\ intensity (black diamonds) and the bottom panels the temporal evolution of B$_{\rm LOS}$ for the photosphere (blue crosses) and the chromosphere (black diamonds) with the fits from Eq.~\ref{eq1} shown with dashed lines. It can be seen that the magnetic field evolutions of the photosphere and the chromosphere are seemingly independent, sometimes showing steps of opposite signs, sometimes showing larger steps in one or the other layer. The changes are permanent and persist even when the chromospheric intensity returns to its original level. Most of the large ($>$300 Mx cm$^{-2}$) chromospheric steps occur right at flare onset (17:45 UT), but later ones ($\sim$17:50) can also be observed. A difference up to 3 minutes can be observed between the timing of photospheric and chromospheric changes.

The AIA emission nearly always starts with an increase in the 1600 and 211 \AA\ passbands, while the other passbands follow with variable delays (seconds to $\sim$10 minutes). Photospheric changes of B$_{\rm LOS}$ seem co-temporal to the increase of AIA emission (within one minute), while chromospheric changes in some cases appear after the AIA emission and the Ca 8542 \AA\ emission peaked (e.g.\,bottom right panels in Fig.~\ref{lightcurves}). There do not seem to be any magnetic field changes \textit{before} the intensity increases in AIA or IBIS. In conclusion, the intensity increase in Ca 8542 \AA\ and AIA passbands and the timing of B$_{\rm LOS}$ changes do not seem to be a direct cause and effect, otherwise the timing would be more consistent, but rather two phenomena occurring near the flare peak time. Also, photospheric and chromospheric changes seem independent of each other. Even though the spatial resolutions of IBIS and HMI are different, a further binning of IBIS to match HMI would not change our conclusions because of the coherent patches of changes in IBIS.

\subsection{Geometry of changes of B$_{\rm LOS}$}

It has been observed that loops contract during and after flares \citep[e.g.,][]{svestkaetal1987,forbesacton1996}, usually on timescales of several minutes, with variable velocities from few km s$^{-1}$ to few hundred km s$^{-1}$ \citep{sunetal2012,liuetal2013,simoesetal2013}. These contractions were linked to the coronal implosion model \citep{hudson2000}, which predicts a decrease of magnetic energy in the corona after a flare. The energy decrease could manifest as contracted, smaller loops and implicitly, the magnetic field at lower atmospheric heights should change by becoming more horizontal. This tilt is consistent to many previous photospheric observations \citep[e.g.][]{petriesudol2010}. Here we investigate if our observed chromospheric changes are compatible to this scenario. The advantage of the chromospheric changes is that they can be visibly attributed to loop footpoints, making it easier to estimate the orientation of the loops.

From our measurements, we identify a patch of pixels in the south-western footpoint which approximately changed from --700 to --400 Mx cm$^{-2}$. Assuming that the field strength remains constant, the tilt to change \mbox{--700} to --400 Mx cm$^{-2}$ depends only on the actual field strength and ranges from 55$^\circ$ (if the initial field was oriented in LOS and thus --700 G) to 12$^\circ$ for an assumed field strength of -1500 G. From chromospheric images (cf. Fig.~\ref{loc1}c) and assuming that loops more or less trace field lines, we can estimate that the field around the large patches of changes was oriented more or less radially. If the field was oriented radially and we measured --700 Mx cm$^{-2}$, this would correspond to an actual field strength of --700/$\mu$ = --875 G and thus a tilt angle of 26$^\circ$ (acos(700/875)-acos(400/875)) would be required. Similarly, for the north-eastern footpoint, a change from 800 to 1000 Mx cm$^{-2}$ was observed, corresponding to a tilt of 37$^\circ$ (for B=1000 G and thus in LOS in the end) to 10$^\circ$ for B=1500 G. B=1500 G was chosen as upper limit for the chromosphere because there is no large umbra in our chromospheric FOV. Again, a LOS field strength of 1000 Mx cm$^{-2}$ corresponds to a radially oriented field of 1250 G, in which case the tilt angle necessary for a change from 800 to 1000 Mx cm$^{-2}$ is 13$^\circ$. Future vector magnetic observations will need to be used to verify the actual tilt angles.

\begin{figure}[tb] 
  \centering 
   \includegraphics[width=.49\textwidth]{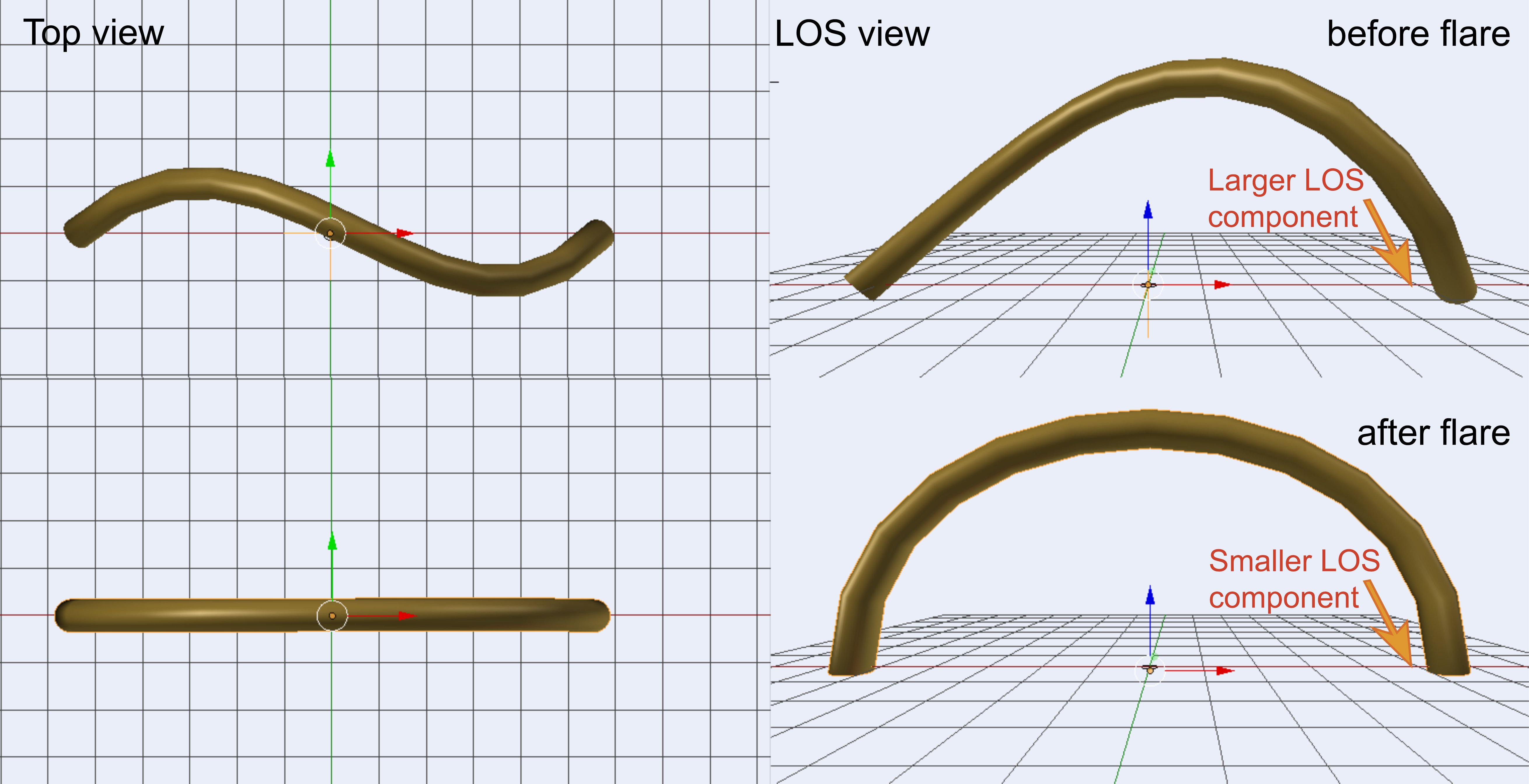}
   \caption{Possible configuration of loops that may explain a decrease in LOS field in the more western (right) and an increase in the eastern (left) footpoint. The assumption here is that the loop untwists during the flare, not requiring any size changes.}
        \label{fluxrope}
  \end{figure}

Because we only measure B$_{\rm LOS}$ and not the vector magnetic field, we are not sensitive to rotations of B along the line-of-sight direction. The flare occurred around coordinates [520, 260]\arcsec, which corresponds to a heliocentric angle of $\theta\approx$ 37$^\circ$, or more commonly $\cos(\theta)=\mu=0.8$. Fig.~\ref{inclination} shows the geometry of the reconnected loops as seen by AIA 131 10 min after flare start. Two loop geometries that connect opposite polarities, locations near both HXR footpoints, and two patches of magnetic field changes were drawn as black arcs. The white arrow points to disk center. The schematic drawing on the right shows these arcs with their footpoint coordinates and the observed changes labeled. The LOS direction is only approximate because projection effects were omitted for simplicity. As long as the north-eastern footpoint is closer to disk center than the south-western footpoint, as in our observation, a loop contraction will lead to a smaller LOS component in the north-eastern footpoint and a larger LOS component in the south-western footpoint. This is the opposite to the field changes we observed. Our observations therefore indicate either increasing loops during the flare or a different, more complex geometry, or possibly newly formed loops that previously were invisible. A possibility to explain the observations is depicted in Fig.~\ref{fluxrope}. If the loops are twisted before the flare and transform into an energetically favorable state without twist after the flare, this may explain the observed LOS changes at their footpoints. But without knowing the exact geometry, we cannot estimate which of the tilt angles we calculated above are most likely. Another problem when trying to attribute the magnetic field changes to loop shrinkages are the timescales. The magnetic field changes are abrupt and often occur within a minute, and no shrinking loop observation to date reported such timescales. Excluding observations of slow loop shrinkages during the post-flare phase and focusing on the fastest example that we found \citep{sunetal2012}, the shrinkage still took about 3 minutes. This timescale may be compatible to some of our slower magnetic field changes, but not to the most abrupt ones (e.g. points in the large southern patch of chromospheric changes, see Fig.~\ref{lightcurves}). A twisting scenario would not require any shrinking or increases. This scenario may be compatible to the implosion model of \citet{hudson2000}, as the downward forces may simply untwist the loops. A reduction in twist accompanied by magnetic field changes has also been speculated for the active region analyzed by \citet{sunetal2012} based on non-linear force-free field extrapolations.

\section{Discussion and Conclusions}

Our findings can be summarized as follows:
\begin{itemize}
\item  Changes of the chromospheric B$_{\rm LOS}$ were found during the X1 flare on 2014-03-29. They are stronger (maximum value 640 Mx cm$^{-2}$) and more extended than photospheric changes.
\item While photospheric changes are located near the polarity inversion line, chromospheric changes seem to predominantly occur near footpoints of coronal loops. In $\sim$60\% of the pixels and in coherent areas, the absolute value of B$_{\rm LOS}$ decreases.
\item The changes are generally near (few arcsec), but not perfectly co-spatial with HXR emission or a small sunquake. Flare-related enhanced AIA emission occurs in nearly all locations that show changes of the magnetic field. The reverse is not valid: AIA emission occurs in many locations without any corresponding change of B$_{\rm LOS}$.
\item The change of B$_{\rm LOS}$ in the photosphere and in the chromosphere is seemingly independent, showing differences in timing ($<$3 minutes), sign, and size.
\item It seems that the timing of enhanced chromospheric and coronal emission is not directly related to the timing of the B$_{\rm LOS}$ change. While the change never occurs before the AIA intensity increase, its delay is variable. Photospheric changes usually occur close to the increase in AIA 1600 and 211 emission. Chromospheric changes may even appear a few minutes after AIA emission peaked.
\item A simple geometrical model of contracting loops does not agree with the observed changes (decrease of $\left|{\rm B}_{\rm LOS}\right|$ in south-western footpoint, increase in north-eastern footpoint). The observations are compatible with increasing loop sizes during the flare or more complex geometries, for example untwisting loops.
\end{itemize}

From previous investigations, we additionally know that the changes are not correlated to sites of enhanced continuum emission \citep{kleintetal2016}, or sites of linear polarization in He 10830 \AA\ \citep{judgeetal2015}. One thing that remains unclear is why AIA 211 shows one of the first responses in pixels with magnetic field changes. For flare emission in general, also for this flare in pixels without B changes, emission in AIA 211 may be delayed compared to the other passbands. The AIA 211 passband has overlaps with many other AIA passbands and therefore, the observed time difference is not yet explained.

Our conclusions about photospheric changes generally agree with previous work. For example, \citet{johnstoneetal2012} similarly found brightenings in 1600 \AA\ where field changes occurred, although the timing had a larger offset in their case (4-9 min before B-change), while it was co-temporal in our case. Our reported size of photospheric changes ($<$320 Mx cm$^{-2}$) is also similar to previous studies \citep{sudolharvey2005, petriesudol2010}. 

This is the first confirmed and quantitative report of chromospheric magnetic field changes during a flare. Their behavior is independent of the photospheric changes and it is unclear why.  The chromospheric super-penumbra or canopy structures around sunspots do not have corresponding photospheric structures and tilts of field lines would therefore only be visible in the chromosphere, which may be an explanation for some differences. Only if the field were vertical and forming large coronal loops, one could expect photospheric and chromospheric changes in the same locations. But it remains unclear why the photospheric field changes significantly and what geometry may be favorable for such changes.

How can our observations be understood in the context of flare models, especially that of coronal implosions \citep{hudson2000}? These models predict that magnetic fields become more horizontal. Statistically, this agrees with our observations of more decreases than increases of B$_{\rm LOS}$, considering that our active region was at heliocentric angle $\mu$=0.8. Such a behavior has already been found in previous observations \citep[e.g.][]{petriesudol2010}. Our new result is that the fraction of decreases/increases is similar for the photosphere and the chromosphere, even though both layers show different areas of changes. But a simple model of contracting loops fails because one should expect the footpoints to show opposite directions of changes than what was observed. Could we be seeing a special case of loops that increase? Considering that a low-lying filament erupted, it may have stretched the field lines upwards. But a more simple explanation may include loops that untwist without any need for shrinkage/increases. A future study should focus on deriving the Lorentz force change taking the chromospheric observations into account. A downward directed change of the Lorentz force has been reported previously \citep[e.g.][]{wangliuwang2012,petrie2013}, but considering that larger areas change in the chromosphere, it would be worthwhile to calculate where the energy is dissipated and why it does not induce measurable changes in parts of the photosphere. The Lorentz force is one of the possible drivers for sunquakes and magnetic jerks, dubbed ``McClymont'' jerks \citep{hudsonetal2008}, may play a role in sunquake generation. It remains unclear why only a tiny sunquake was observed for this flare and not even near the largest magnetic field changes.

When a change of chromospheric magnetic field is observed, could it simply be that the optical depth changes during the flare and other layers with different field strength become visible, only making it appear as if the magnetic field changed? This seems unlikely because intensities return to pre-flare values, while the field changes are permanent. But it is possible that the chromospheric structure dynamically evolves in a way that new structures with different magnetic field strengths move into the line-of-sight. This may be an alternate explanation for apparent magnetic field changes and very difficult to verify because the chromospheric structures visibly change during the flare.

Several questions remain open. If changes arise due to the tilting of field lines that form coronal loops, chromospheric pixels would be expected to show slightly larger changes than those in the photosphere, but the variable signs and strengths of changes in both layers rule out such a simple picture. It is also yet unclear which process causes the field to change and in which locations, because observations do not show any direct relation of the magnetic field changes to any flare related phenomena (white light, HXR, ribbons). Further multi-height polarimetric observations are needed to clarify the magnetic restructuring during flares.


\acknowledgments
I thank Phil Judge, Hugh Hudson, S\"am Krucker, and Sebastian Castellanos Duran for valuable suggestions and discussions and Alina Donea for the sunquake map and explanations. This work was supported by a Marie Curie Fellowship and the NASA grants NNX13AI63G / NNX14AQ31G. The observations were obtained together with Alberto Sainz Dalda, Kevin Reardon and Paul Higgins at the NSO, which is operated by the Association of Universities for Research in Astronomy, Inc. (AURA), for the National Science Foundation. IBIS is a project of INAF/OAA with additional contributions from University of Florence and Rome and NSO.
This work has benefited from discussions during an ISSI meeting about high-resolution chromospheric observations. I am very grateful to all the observers of the DST without whom these observations would not have been achieved and dedicate this article to the memory of Mike Bradford.

\bibliographystyle{apj}
\bibliography{journals,ibisflare}

\end{document}